\documentstyle[12pt,epsfig]{article}

\newlength{\dinwidth}
\newlength{\dinmargin}
\def\lapproxeq{\lower .7ex\hbox{$\;\stackrel{\textstyle
<}{\sim}\;$}}
\def\gapproxeq{\lower .7ex\hbox{$\;\stackrel{\textstyle
>}{\sim}\;$}}

\def\beq{\begin{equation}}
\def\eeq{\end{equation}}
\def\bea{\begin{eqnarray}}
\def\eea{\end{eqnarray}}

\def\bb{b\bar{b}}
\def\tt{t\bar{t}}

\def\GeV{\rm GeV}
\def\msb{\overline{\rm MS}}

\begin{document}
\begin{flushright}
Cavendish-HEP-2007/12 \\

\end{flushright}

\vspace*{0.5cm}

\begin{center}
{\Large \bf Parton Distributions for LO Generators}

\vspace*{1cm}
\textsc{A. Sherstnev$^a$ and R.S. Thorne$^{b,}$\footnote{Royal 
Society University Research Fellow}} \\

\vspace*{0.5cm} $^a$ Cavendish Laboratory, University of Cambridge, 
\\ JJ Thomson Avenue, Cambridge, CB3 0HE, UK\\ and Moscow State University, Moscow, Russia (on leave)\\
$^b$ Department of Physics and Astronomy, \\
University College London, WC1E 6BT, UK
\vspace*{0.5cm}

\end{center}

\vspace*{0.5cm}

\begin{abstract}
We present a study of the results obtained by combining LO partonic matrix 
elements with either LO or NLO partons distributions. These are compared 
to the {\em best} prediction using NLO for both matrix elements and parton 
distributions. The aim is to determine which parton distributions are most 
appropriate to use  in those cases where only LO matrix elements are 
available, 
e.g. as in many Monte Carlo generators. Both LO and NLO parton distributions 
have flaws, sometimes serious, for some processes, so a modified {\em optimal} 
LO set is suggested. We investigate a wide variety of process, and the new modified 
LO* pdf works at least as well as, and often better than, both LO and NLO pdfs 
in nearly all cases. The LO* pdf set is now available in the LHAPDF package~\cite{Whalley:2005nh}.
\end{abstract}

\begin{figure}
\begin{center}
\centerline{
\epsfxsize=0.7\textwidth\epsfbox{{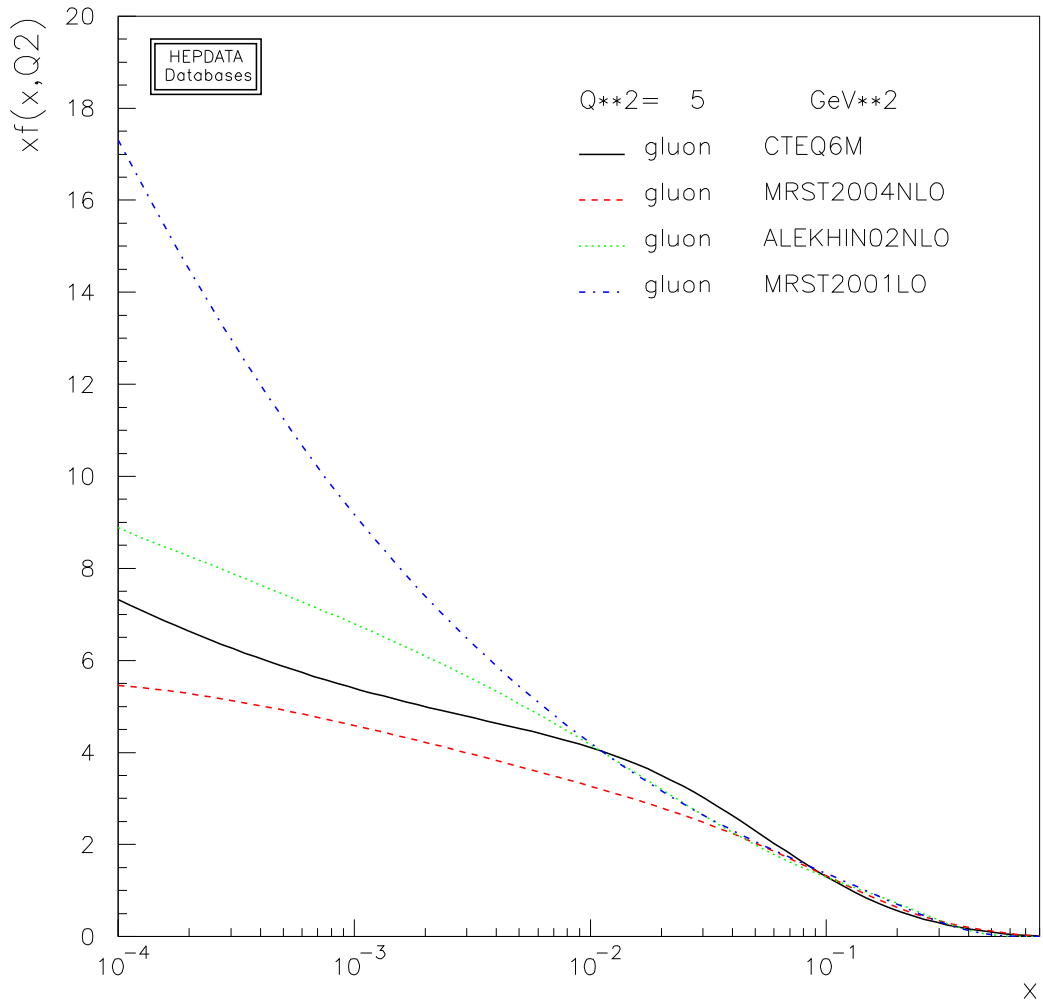}}
}
\vspace{0.3cm}
\caption{ 
A comparison of various gluon distributions at NLO and a gluon distribution 
at LO. There is significant variation in the NLO distributions, but the LO 
distribution is a qualitatively different shape from all of them. 
}
\label{locomp}
\end{center}
\end{figure}

\section{Introduction}
It has long been known that for certain regions of $x$ there can be big 
differences between parton distributions extracted at different orders of 
perturbative QCD (see for example \cite{Thorne:2006wq} for a discussion). 
Moreover these 
differences between LO, NLO, and NNLO partons are much more significant 
than those between parton distributions extracted by various groups. This 
happens due to important missing higher order corrections in the splitting 
functions which determine the parton evolution, and in the cross-sections 
which govern their extraction by comparison to experimental data (mainly 
from structure functions in deep inelastic scattering). In particular, use 
of parton distributions of the wrong order can lead to incorrect 
conclusions on the size of the small-$x$ gluon, shown in Fig.~\ref{locomp}, 
and consequently on the importance of corrections due to shadowing, 
saturation {\it etc.} at small $x$. 

This large change in the size of parton distributions as one changes 
perturbative order means that the combination of the respective order of 
the parton distribution function (pdf) and the accompanying matrix element 
is an important issue when calculating the cross-section for any physical 
process~\cite{Thorne:2007jc}. Traditionally, LO pdfs are thought to be the 
best choice for use with LO matrix elements, such as those available in many 
LO Monte Carlo programs, though it has been recognised that all such results 
should be treated with care. However, recently an alternative viewpoint has 
appeared, and it has been suggested that NLO pdfs may be more appropriate~\cite{Campbell:2006wx}.
The argument is that NLO cross-section corrections are often small, and the 
main change in the total cross-section in going from LO to NLO is due to the 
pdfs. Indeed, there has already been an investigation of the use of NLO pdfs 
for the underlying event~\cite{Albrow:2006rt} at the Tevatron. There is a big 
difference in the results when using CTEQ6L and CTEQ6.1M pdfs~\cite{Pumplin:2002vw} 
due to the changes in the gluon, though agreement can be reached by 
significant retuning. However, this retuning will potentially affect 
predictions for other quantities. 

In this article we investigate the differences in predictions obtained for a 
variety of physical quantities combining different  pdfs with LO matrix 
elements. In each case we assume NLO pdfs combined with NLO matrix elements represent 
the best prediction -- the {\it truth}.\footnote{Of course, the full NNLO 
calculation is known for some processes, but we will avoid this further 
complication in this article.} The different processes chosen rely on a 
variety of input parton  distributions and on various ranges of $x$ and $Q^2$ 
in order for us to try to make our conclusions as all-encompassing as possible. 
We interpret the features of the results noting that there are significant 
faults if one uses exclusively either LO or NLO pdfs. We hence attempt to 
minimise this problem, and investigate how a best set of pdfs for use with 
LO matrix elements may be obtained.

\section{Parton Distributions at Different Orders}

\begin{figure}
\begin{center}
\centerline{
\epsfxsize=0.45\textwidth\epsfbox{{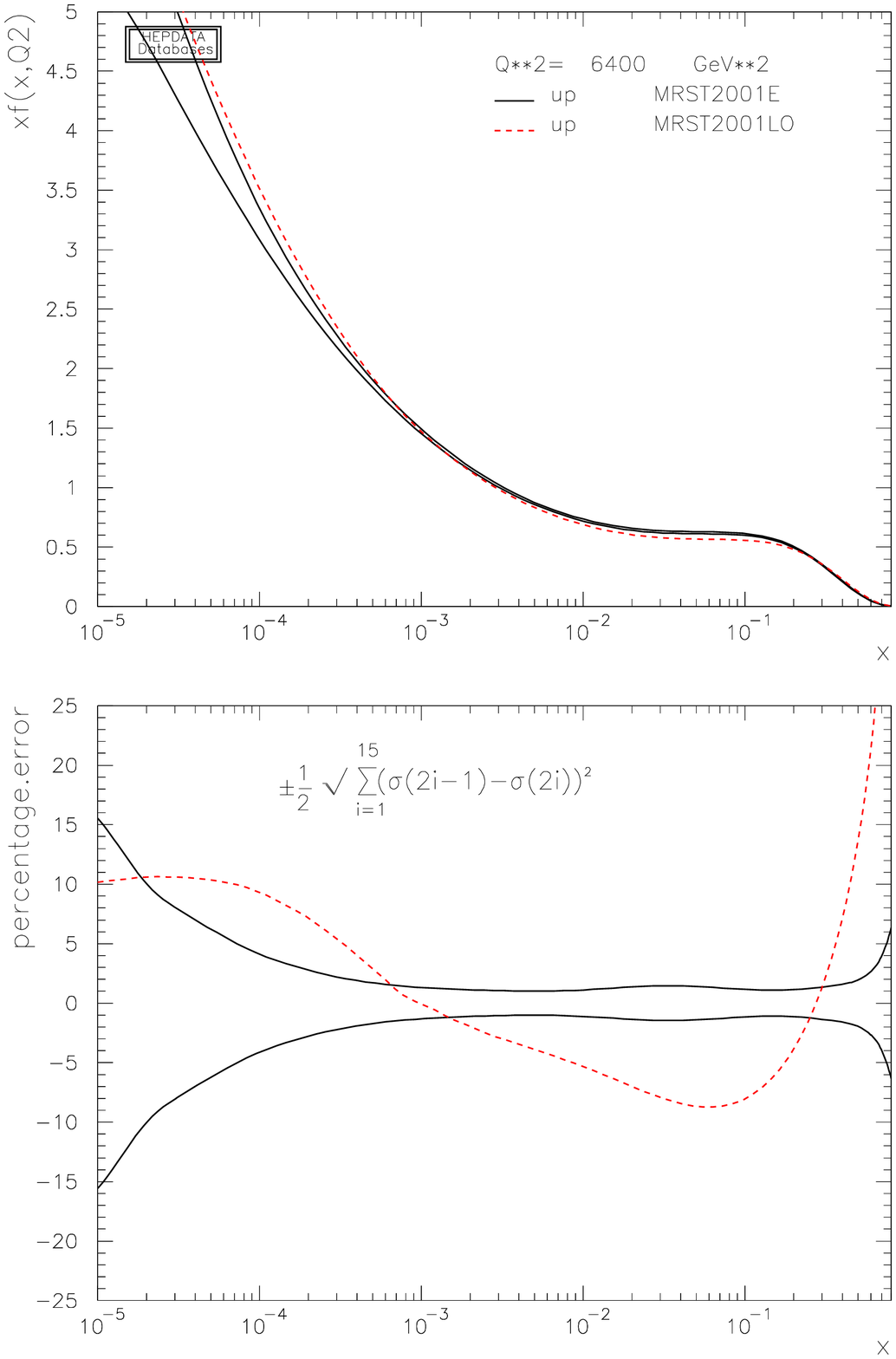}}
\hspace{0.7cm}
\epsfxsize=0.45\textwidth\epsfbox{{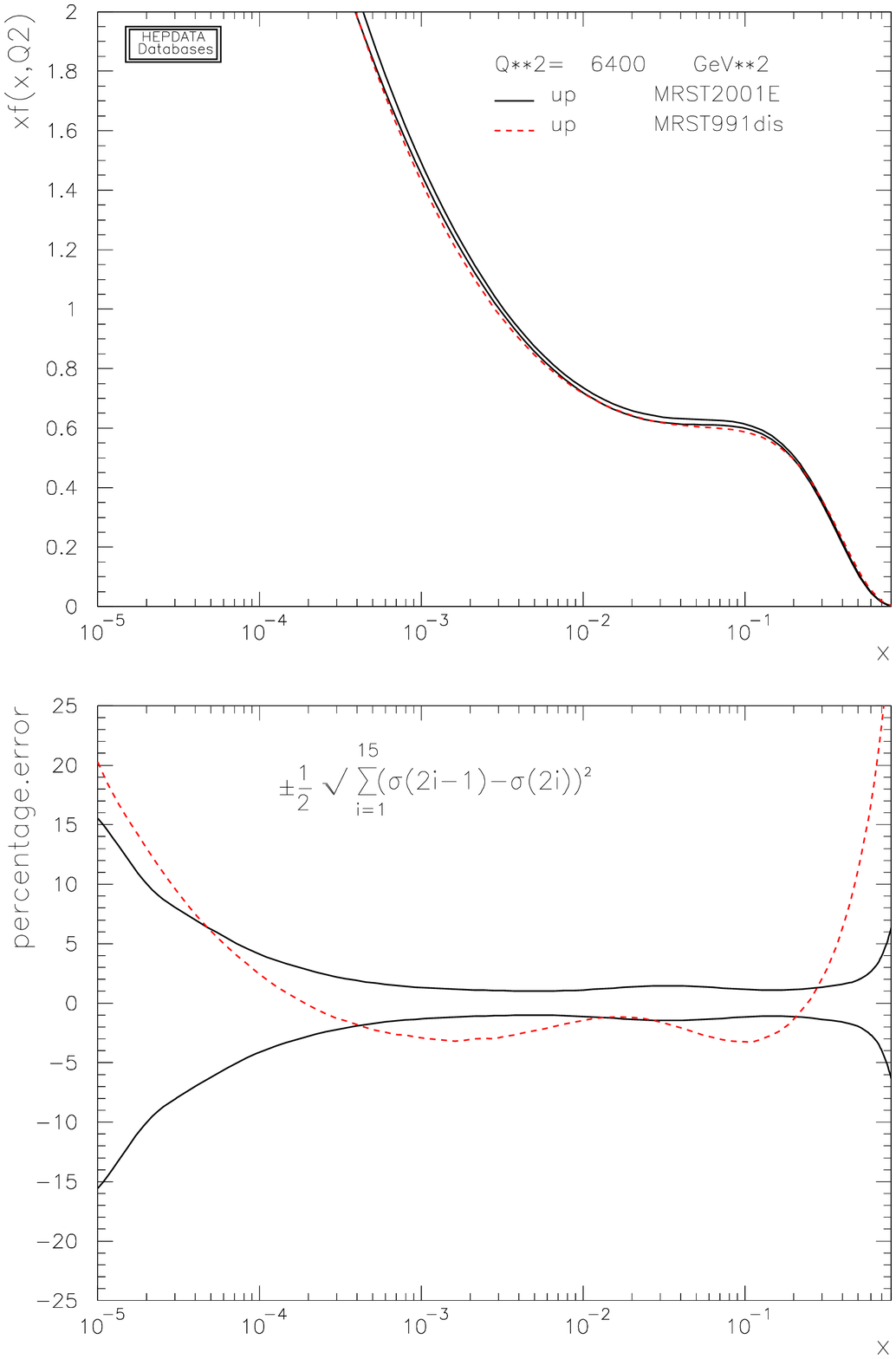}}
}
\vspace{0.3cm}
\caption{ 
The comparison of the up quark parton distribution at LO with that at NLO 
in the $\msb$ factorization scheme (left) and the comparison between the 
NLO $\msb$-scheme and DIS-scheme up quark distribution (right), both at at 
$Q^2=6400\GeV^2$. The upper plots show the uncertainty bands of the NLO $\msb$
distribution and the lower plots the percentage uncertainty. In both cases 
this is compared to the central LO or DIS scheme distribution.  
}
\label{mrstlonlo}
\end{center}
\end{figure}

Let us briefly explain the reasons for the origins of the differences between 
the parton distributions at different perturbative order. The qualitative 
difference between the LO and NLO gluon distribution is shown in 
Fig.~\ref{locomp}. Clearly the LO gluon is much larger at small $x$ than any 
NLO gluon at $Q^2=5\GeV^2$. The evolution of the gluon at LO and NLO is quite 
similar, so at larger $Q^2$ the relative difference is smaller, but always 
remains significant. At $Q^2=1-2\GeV^2$ the difference is even more 
marked, with NLO gluon distributions becoming valence-like, or even negative 
at very small $x$, while the LO distribution remains large. This difference 
in the gluon distributions is a consequence of quark evolution, rather than 
gluon evolution. The small-$x$ gluon is determined by $dF_2/d\ln Q^2$, which 
is directly related to the $Q^2$ evolution of the quark distributions. The 
quark-gluon splitting function  $P_{qg}$ is finite at small $x$ at LO, but 
develops a small-$x$ divergence at NLO (and further $\ln(1/x)$ enhancements 
at higher orders), so the small $x$ gluon needs to be much bigger at LO in 
order to fit structure function evolution. 
 
There are also significant differences between the LO and NLO quark 
distributions, as shown in the left of Fig.~\ref{mrstlonlo}. Before 
analysing these in detail it is important to realise that beyond LO the 
parton distributions are factorization scheme dependent. We 
conventionally take NLO to mean ``NLO in the $\msb$ factorization scheme'', 
and indeed this is the NLO distribution used in the left of 
Fig.~\ref{mrstlonlo}. In order to illustrate the difference between orders we 
also show the comparison of the $\msb$-scheme distribution at NLO and the 
DIS-scheme distribution in the right of Fig.~\ref{mrstlonlo}. In the 
DIS-scheme the NLO coefficient functions relating parton distributions to 
structure functions are zero. Hence, if the LO parton distributions and NLO 
distributions produced identical structure functions, the right and left plots 
in Fig.~\ref{mrstlonlo} should be the same. Indeed, there are obvious 
similarities, both LO and NLO DIS-scheme quark distributions being larger than NLO 
$\msb$-scheme distributions at very high $x$ and very low $x$. This is due 
to the effect of the NLO coefficient functions in the $\msb$-scheme. Most particularly 
the quark coefficient functions for structure functions in the $\msb$ scheme have 
$\ln(1-x)$ enhancements at higher perturbative order, and the high-$x$ quarks 
are smaller as the order increases. However, the LO quark distribution is 
depleted compared to the NLO DIS-scheme quark distribution for 
$0.001\leq x \leq 0.1$. And indeed, the LO fit to structure function data is 
poor in this region. 

Hence, while LO parton distributions are more similar to NLO DIS-scheme 
distributions than NLO $\msb$-scheme distributions (the gluons in 
DIS scheme and $\msb$ scheme are similar except at very high $x$), there are 
some qualitative differences between LO and NLO in general. The LO gluon is 
much bigger at small $x$, and compared to $\msb$ scheme the LO valence quarks
are much bigger at high $x$. This is then accompanied by a significant 
depletion of the quark distribution for $x$ in the region of 0.01, despite 
the fact that this leads to a poor fit to data. 

\begin{figure}
\begin{center}
\centerline{
\epsfxsize=0.7\textwidth\epsfbox{{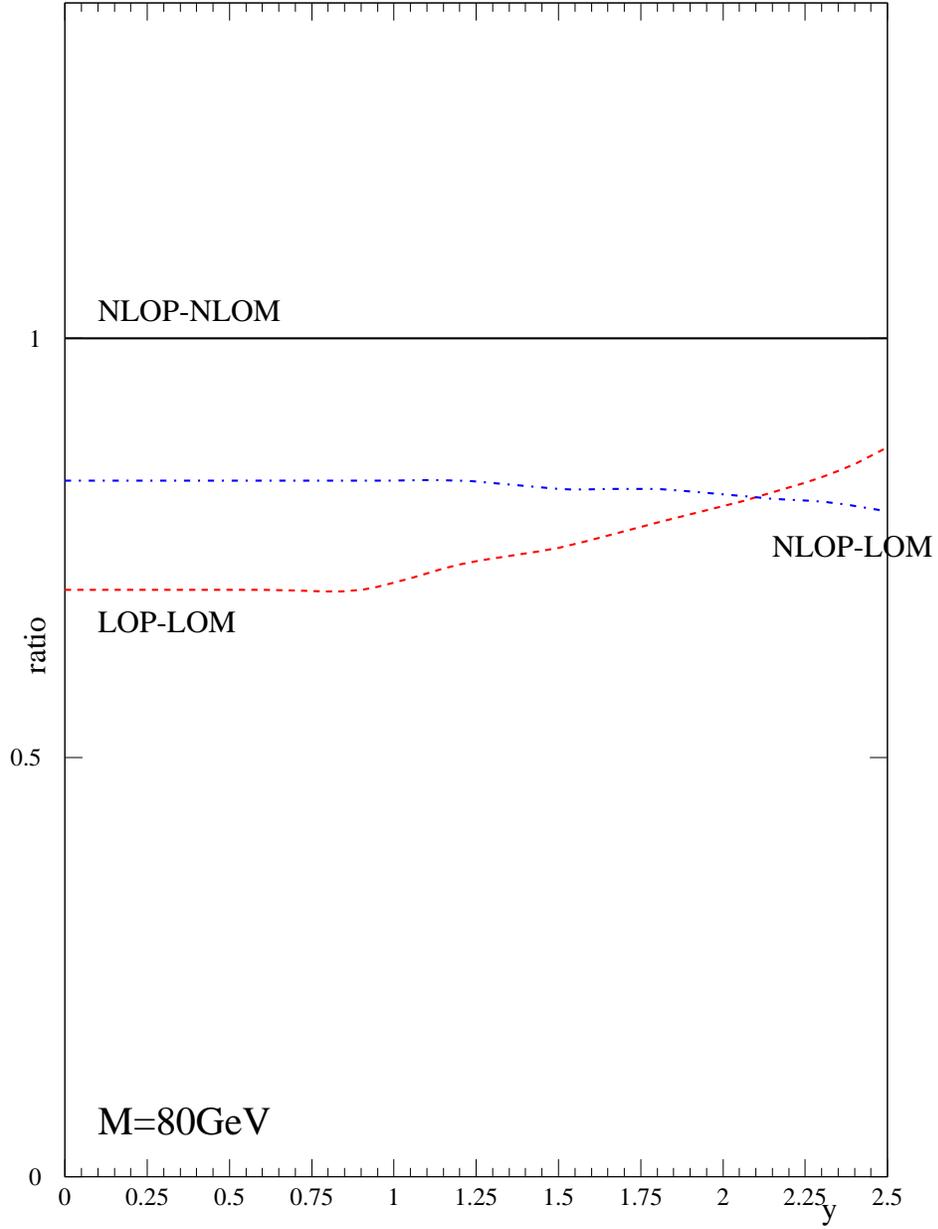}}
}
\vspace{0.3cm}
\caption{ 
Comparison of boson production at the Tevatron using combinations of 
different orders of parton distributions and hard partonic cross-sections. 
}
\label{monterap}
\end{center}
\end{figure}

Let us examine the consequence of these differences when calculating physical
cross-sections. In Fig.~\ref{monterap} we illustrate the ratio of the 
cross-sections produced for Drell-Yan production of a boson with invariant mass
$80\GeV$ at Tevatron energies using NLO $\msb$-scheme parton distributions 
and NLO $\msb$-scheme matrix elements (taken as the reference) \cite{Kubar:1980zv}, 
NLO $\msb$-scheme parton distributions and LO matrix elements and 
LO parton distributions and LO matrix elements.\footnote{Since NLO matrix 
elements are most readily available in $\msb$ scheme, we will take this as 
the default, and henceforth NLO is intended to mean NLO in $\msb$ scheme 
unless explicitly stated otherwise.} In this case for central rapidity, $y=0$, 
both partons have $x=0.04$, and the larger and smaller $x$, $x_1$ and $x_2$ 
are $0.04\exp(y)$ and $0.04\exp(-y)$ respectively. In this case it is the quark 
distributions which are probed. 

Clearly we are generally nearer to the {\it truth} with the LO matrix element 
and NLO parton distribution~\cite{Martin:2004ir} than with the LO matrix element and 
LO parton distributions~\cite{Martin:2002dr}. However, this is always too small, 
since the NLO correction to the matrix element is positive. It is relatively 
$y$-independent, but there is the beginning of extra suppression at the 
highest $y$ for the PDF[NLO]-ME[LO] calculation due to an additional enhancement 
in the NLO matrix element at high $y$, similar to the $\ln(1-x)$ enhancement 
for structure function coefficient functions. The depletion of the LO quark 
distributions for $x\sim 0.04$ leads to the extra suppression in the 
PDF[LO]-ME[LO] calculation.
However, when probing the high-$x$ quarks, the increase in the LO parton 
compensates for the increase in the NLO matrix element, and for $y>2$ this gives 
the more accurate result. However, overall the shape as a function of $y$ is 
much worse using the LO parton distributions than the NLO distributions.  

\begin{figure}
\begin{center}
\centerline{
\epsfxsize=0.7\textwidth\epsfbox{{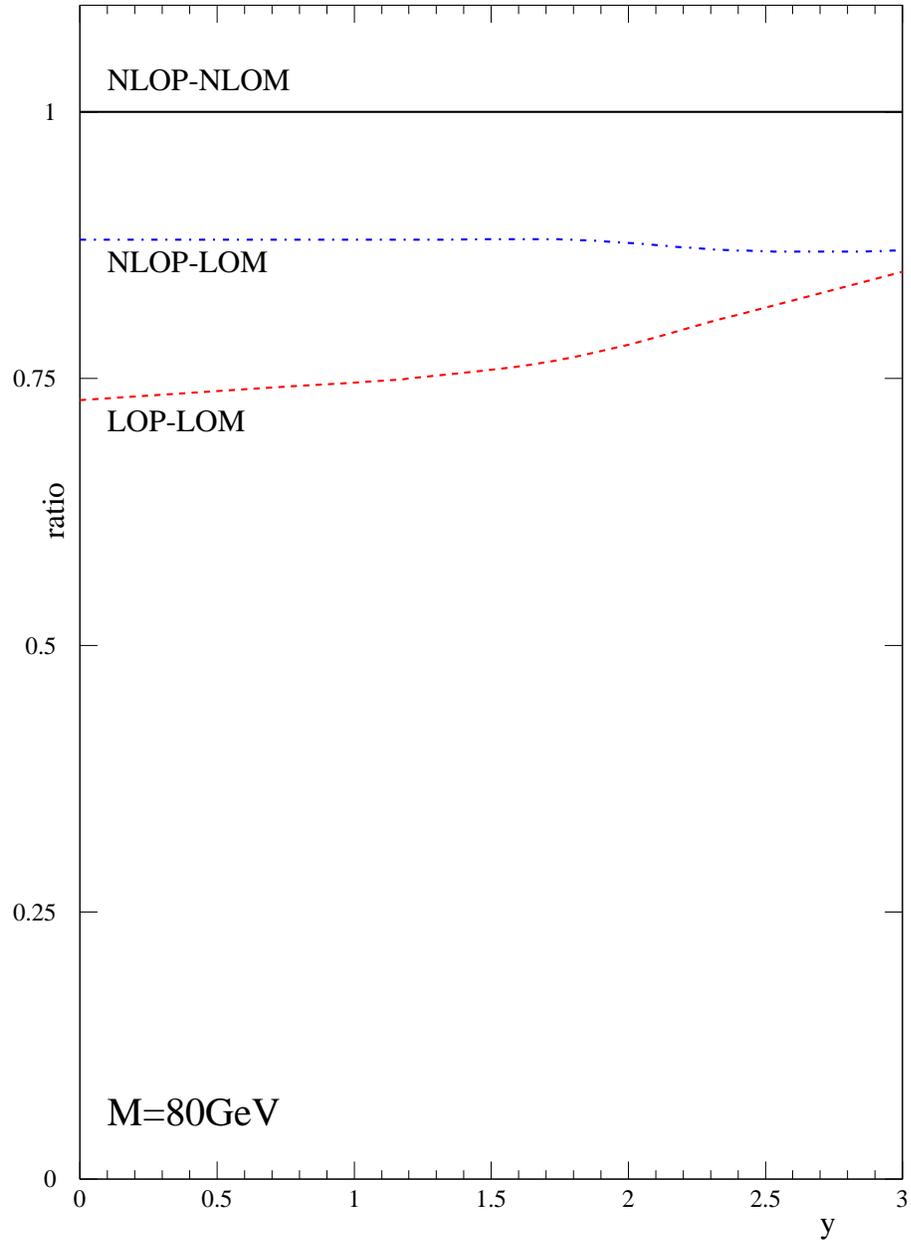}}
}
\vspace{0.3cm}
\caption{ 
The same as Fig.~\ref{monterap}, but at LHC energies. 
}
\label{monteraplhc}
\end{center}
\end{figure} 

We consider the same comparison, but for boson production at the LHC rather 
than the Tevatron, in Fig.~\ref{monteraplhc}. In this case central rapidity
corresponds to $x=0.006$, and we have to go to higher rapidity than at the 
Tevatron to be sensitive to large-$x$ effects, but have much more sensitivity 
to small $x$. Again the PDF[NLO]-ME[LO] calculation is suppressed compared to the 
{\it truth}, but this time there is less change in shape at high rapidity
since even at $y=3$ ($x_1=0.11, x_2=0.0003$) we are barely reaching the 
regime of enhanced NLO partonic cross-sections. As for the Tevatron, the 
PDF[LO]-ME[LO] result is even more suppressed, due to the depletion of quarks, 
though this begins to reduce at the highest $y$, mainly due to increases in 
small-$x$ quarks at LO. The general conclusion is the same as for the Tevatron
-- the NLO pdfs provide a better normalization and a better shape.

\begin{figure}
\begin{center}
\centerline{
\epsfxsize=0.45\textwidth\epsfbox{{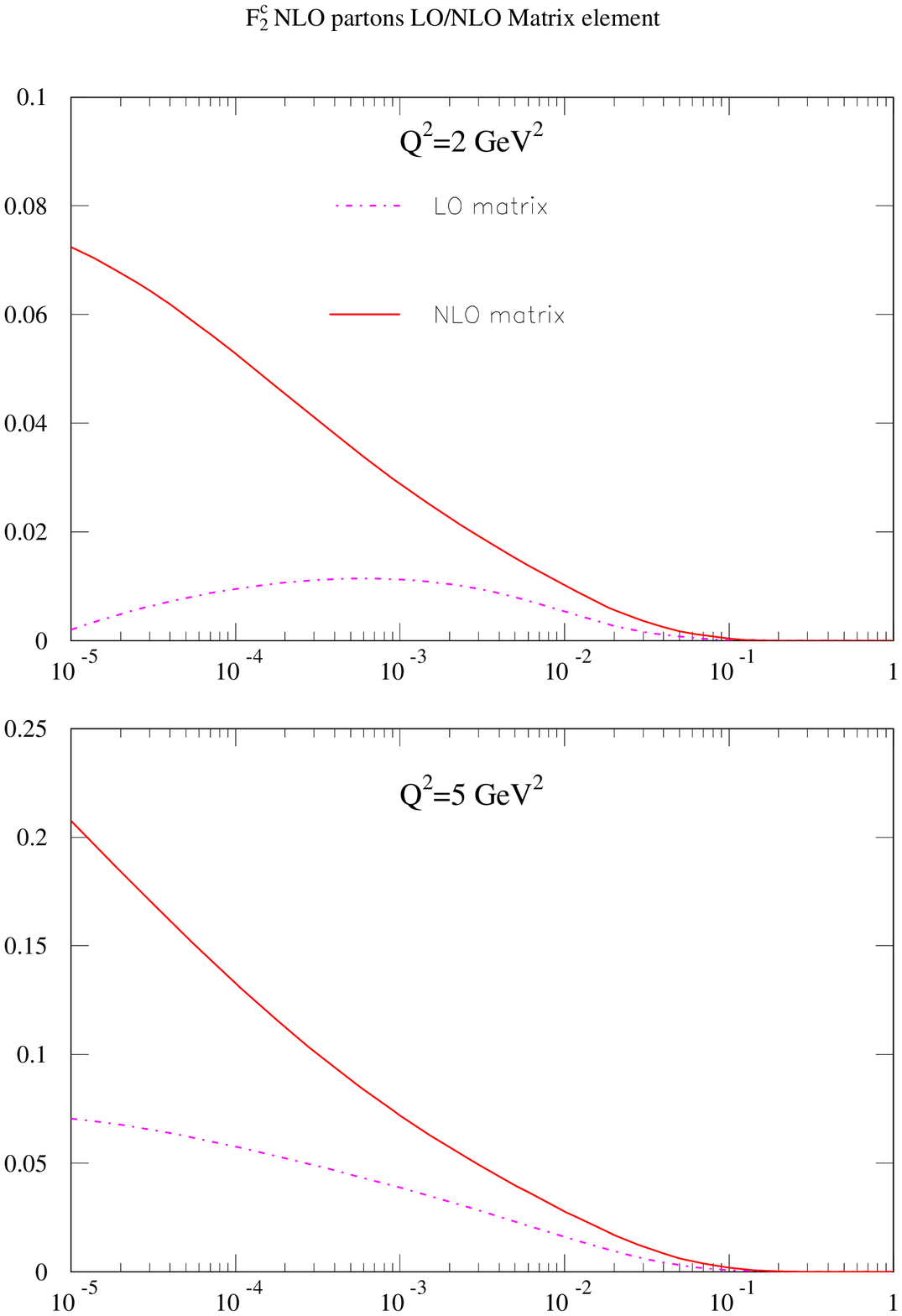}}
\epsfxsize=0.45\textwidth\epsfbox{{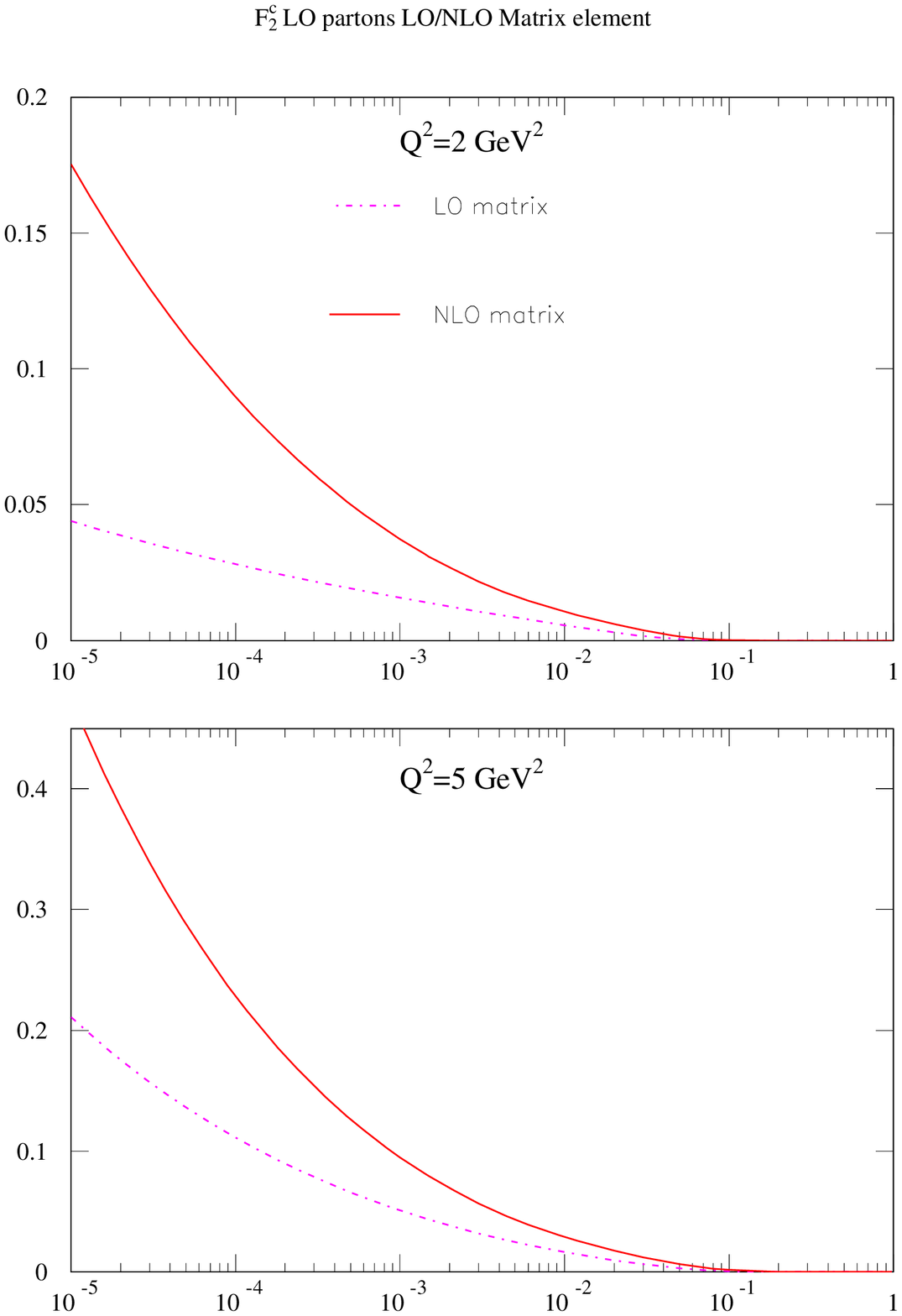}}
}
\vspace{0.3cm}
\caption{
A comparison of charm production at HERA using LO and NLO coefficient 
functions and NLO parton distributions (left), and using LO parton 
distribution functions (right).
}
\label{charmlonlo}
\end{center}
\end{figure} 

\begin{figure}
\begin{center}
\centerline{
\epsfxsize=0.7\textwidth\epsfbox{{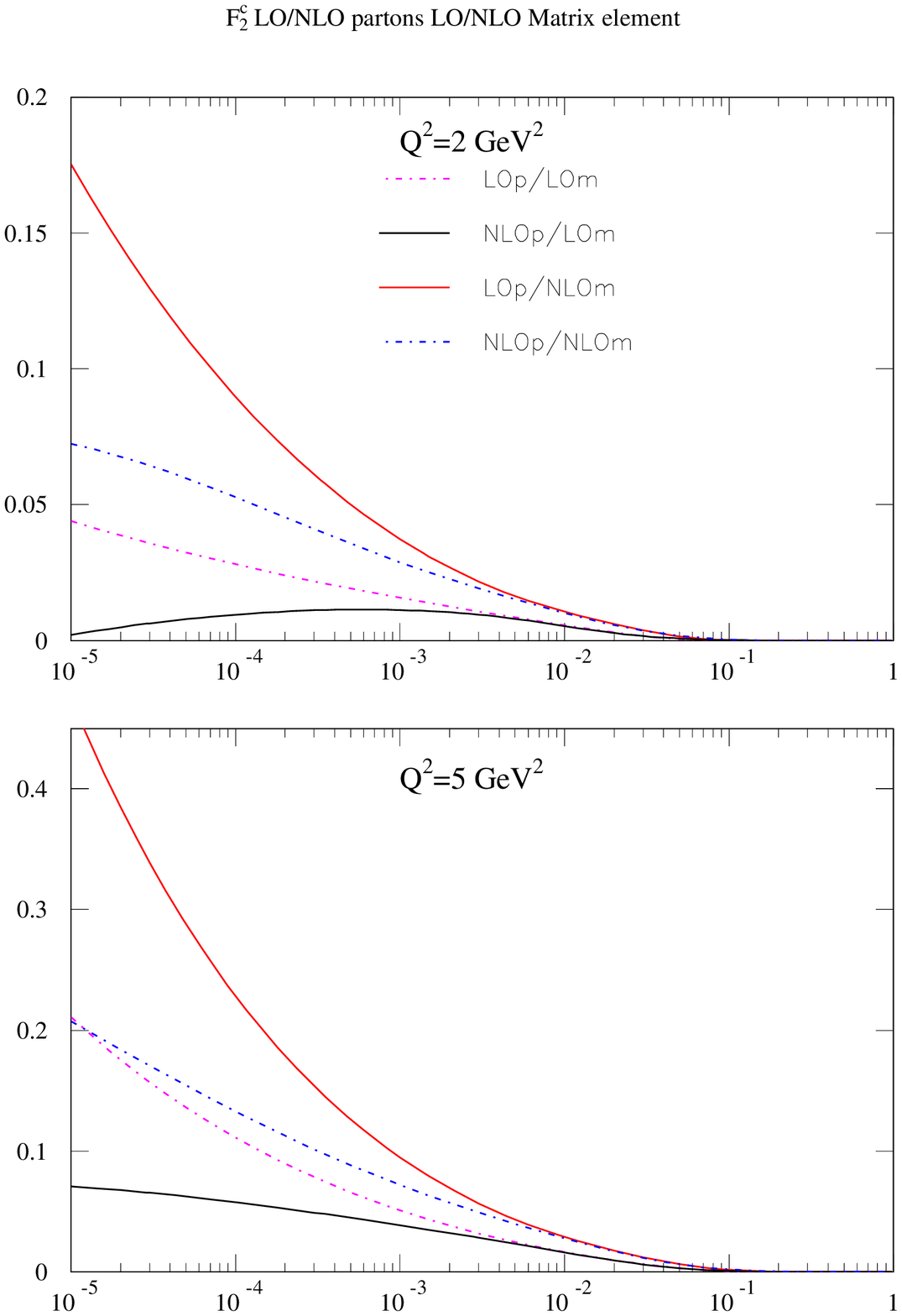}}
}
\vspace{0.3cm}
\caption{ 
A comparison of charm production at HERA using all combinations 
of LO and NLO partons and LO and NLO coefficient functions. 
}
\label{charmlonlob}
\end{center}
\end{figure}

This single quantity suggests that the opinion in~\cite{Campbell:2006wx} is correct, 
i.e. that NLO pdfs are more appropriate if one only has LO matrix elements. 
However, this is a very processes dependent statement. In order to demonstrate 
this we consider a quantity sensitive to the small-$x$ gluon distribution, 
where the difference between the LO and NLO gluon distribution is due to 
missing higher order perturbative corrections. 
Let us consider the production of charm in DIS, i.e. $F^{c\bar c}_{2}(x,Q^2)$ 
where we take all charm as produced in the final state, i.e. we work in the 
fixed-flavour number scheme (FFNS). In this case the NLO coefficient function, 
$C^{c\bar c,(2)}_{2,g}(x,Q^2,m_c^2)$ contains a divergence at small $x$ not 
present at LO, in the same way that the quark-gluon splitting function does, 
the latter being responsible for the large difference between the LO and NLO 
gluon distributions at small $x$. In the left of Fig. \ref{charmlonlo} we see 
the large effect of the NLO coefficient functions \cite{nlocalc}. When using 
NLO partons the LO matrix element result is well below the {\it truth} at low 
scales, and the shape is totally wrong. The structure function follows the 
shape of the gluon distribution at NLO, whereas the small-$x$ divergence in 
the coefficient function provides a charm structure function which rises at 
small $x$ in the same way that the NLO correction to $P_{qg}$ provides a total 
$dF_2/d\ln(Q^2)$ which rises at small $x$. In the right of Fig.~\ref{charmlonlo} 
we see the result using the LO pdfs. In this case $F^{c\bar c}_2(x,Q^2)$ rises 
at small $x$ even with the LO coefficient function, due to the rising gluon. 
Using the NLO coefficient function and the LO pdfs effectively double counts 
the small-$x$ divergence, and the result is much too large and steep. All 
four results are shown together in Fig.~\ref{charmlonlob}. While the LO pdfs 
combined with LO coefficient functions is not a perfect match to the {\it 
truth} -- after all the small-$x$ divergences are not exactly the same in matrix 
element and splitting function -- it is clearly far better than the other two 
approaches which either double count the small-$x$ divergences, or include 
neither. In particular, in this case the NLO pdfs together with the LO matrix 
elements fail badly, providing a clear contradiction to the view 
in~\cite{Campbell:2006wx}. 

Hence, from these two simple examples alone we can make the following 
conclusions. Sometimes it is better to use NLO pdfs if only LO matrix 
elements are known, and one can encounter significant problems with both 
normalization and shape if LO pdfs are used. However, one can be completely 
wrong, particularly at small $x$, by using NLO partons with LO matrix 
elements, due to effective zero-counting of small-$x$ divergences. 
One could finish here and conclude that it is necessary to consider on a case 
by case basis, and in any analysis which uses LO Monte Carlo generators 
sometimes use LO pdfs and sometimes NLO pdfs, depending on the particular 
process under consideration. However, we will continue our investigation 
rather than accepting this pessimistic conclusion and attempt to discover 
whether there are some {\it optimal} partons which have the desirable features
of both LO and NLO pdfs.

\section{Improved LO Parton Distributions}

\begin{figure}
\begin{center}
\centerline{
\epsfxsize=0.7\textwidth\epsfbox{{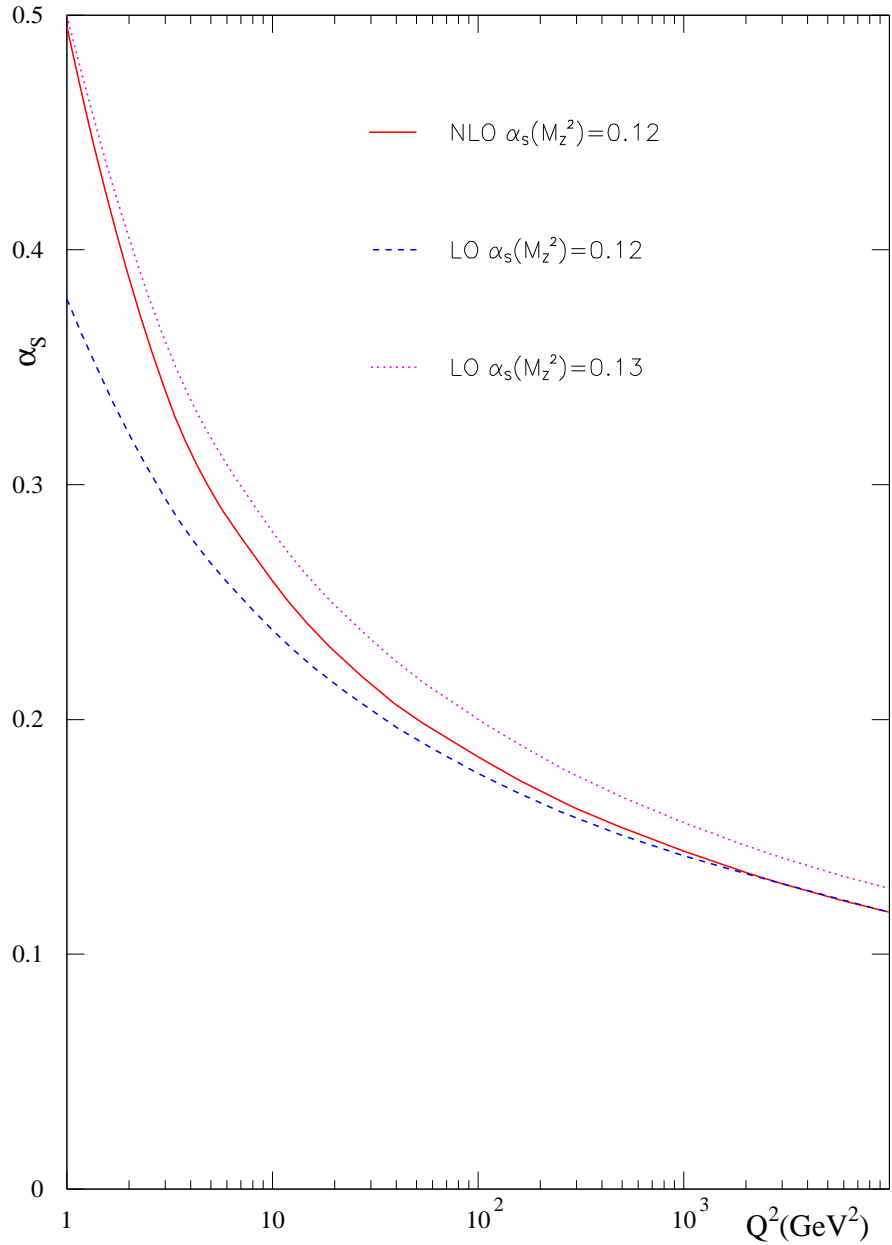}}
}
\vspace{0.3cm}
\caption{ 
Comparison of the LO and NLO definitions of $\alpha_S$. 
}
\label{alphalonlo}
\end{center}
\end{figure}

In order to make progress in finding some {\it optimal} set of pdfs for use 
with LO matrix elements we need to understand fully the differences between 
LO and NLO partons. As seen by comparison with the NLO DIS-scheme partons in 
Fig.~\ref{mrstlonlo}, part of the dip in LO quarks compared to NLO is due to 
extra coefficient function contribution at NLO (particularly for $x \sim 0.1$), 
but it is mostly a problem at LO -- the size and extent of the depletion 
from $x=0.1-0.001$ is reflected in the fit quality. At LO, compared to NLO 
(and higher orders), missing terms in $\ln(1-x)$ and $\ln(1/x)$ in coefficient 
functions and/or evolution lead to a gluon which is much bigger as $x \to 0$ 
and valence quarks which are much larger as $x \to 1$ in order to compensate. 
From the momentum sum rule there are then not enough partons to go around, 
hence the depletion in the quark distributions at moderate to small $x$.

This depletion leads to a bad global fit at LO, particularly for HERA 
structure function data which are very sensitive to quark distributions at 
$x \sim 0.01$. In practice the lack of partons at LO is partially compensated 
by a LO extraction of $\alpha_S(M_Z^2) \sim 0.130$, i.e. the very large 
coupling constant leads to quicker evolution, making up for the higher 
order corrections and/or smaller parton distributions in some regions. This 
results in one obvious modification. It is helpful to use the NLO definition 
of the coupling constant in a LO fit to parton distributions. Because of 
quicker running at NLO, LO and NLO couplings with the same value of 
$\alpha_S(M_Z^2)$ become very different at lower scales where DIS data 
exist, as shown in Fig.~\ref{alphalonlo}. Alternatively, near $Q^2=1\GeV^2$ 
the NLO coupling with $\alpha_S(M_Z^2)=0.120$ is similar to the LO coupling 
with $\alpha_S(M_Z^2)=0.130$. Consequently, the NLO coupling with 
$\alpha_S(M_Z^2)=0.120$ is much bigger than the LO coupling with 
$\alpha_S(M_Z^2)=0.120$, and will do a much better job of fitting the 
low-$Q^2$ structure function data. Hence, the use of the NLO coupling helps 
alleviate the discrepancy between the parton distributions at different 
orders. Indeed, the NLO coupling is already used in some CTEQ LO pdfs and 
in Monte Carlo generators. 

However, even with this modification the LO fit is still poor compared with 
NLO, and the partons are still depleted in some regions. The problems caused 
due to the depletion of partons have led to a suggestion~\cite{Sjostrand} 
that relaxing the momentum sum rule for the input parton distributions 
could make LO partons rather more like NLO partons where they are normally 
too small, while allowing the resulting partons still to be bigger than NLO 
where necessary, i.e the small-$x$ gluon and high-$x$ quarks. 

\begin{figure}
\begin{center}
\centerline{
\epsfxsize=0.45\textwidth\epsfbox{{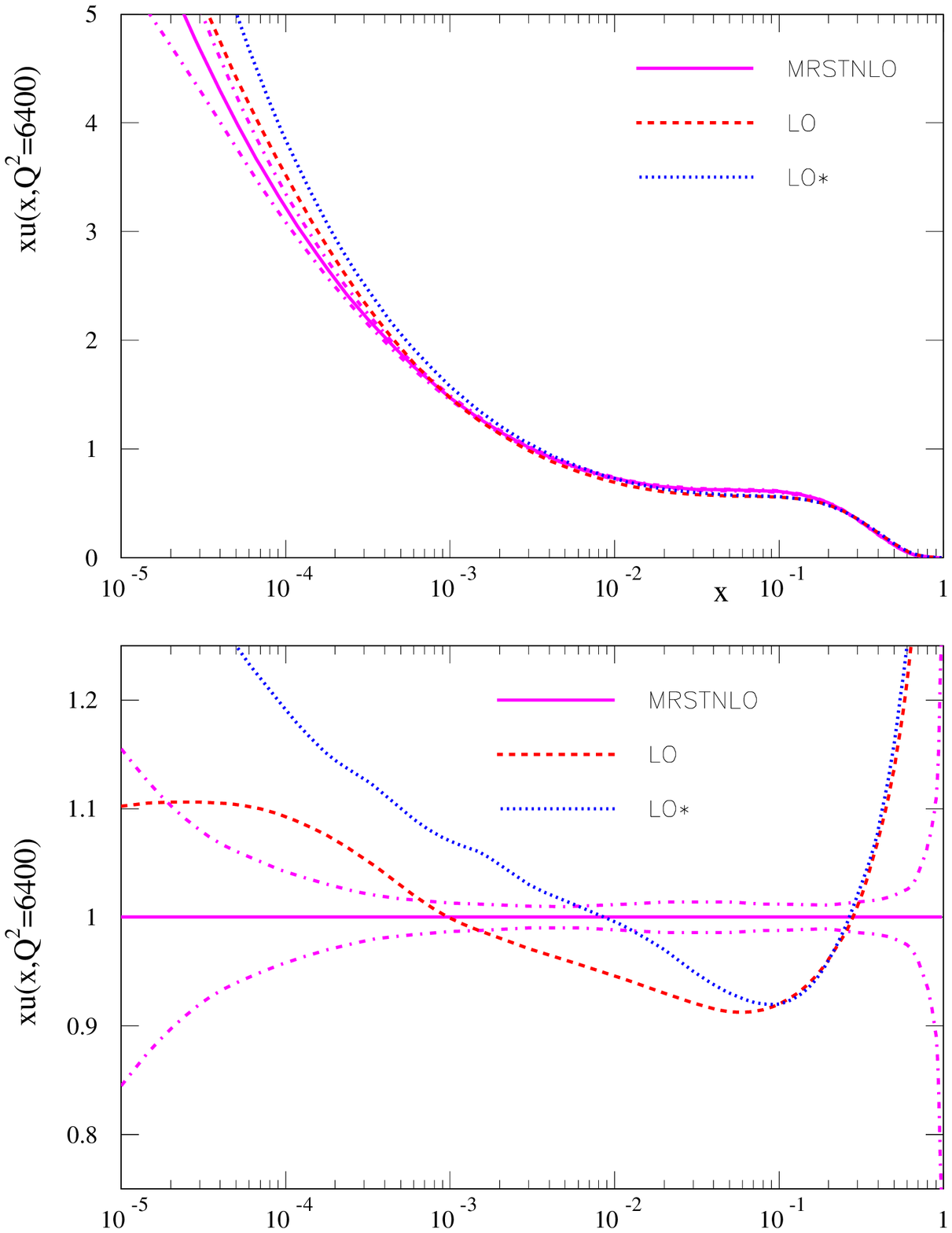}}
\hspace{0.5cm}
\epsfxsize=0.45\textwidth\epsfbox{{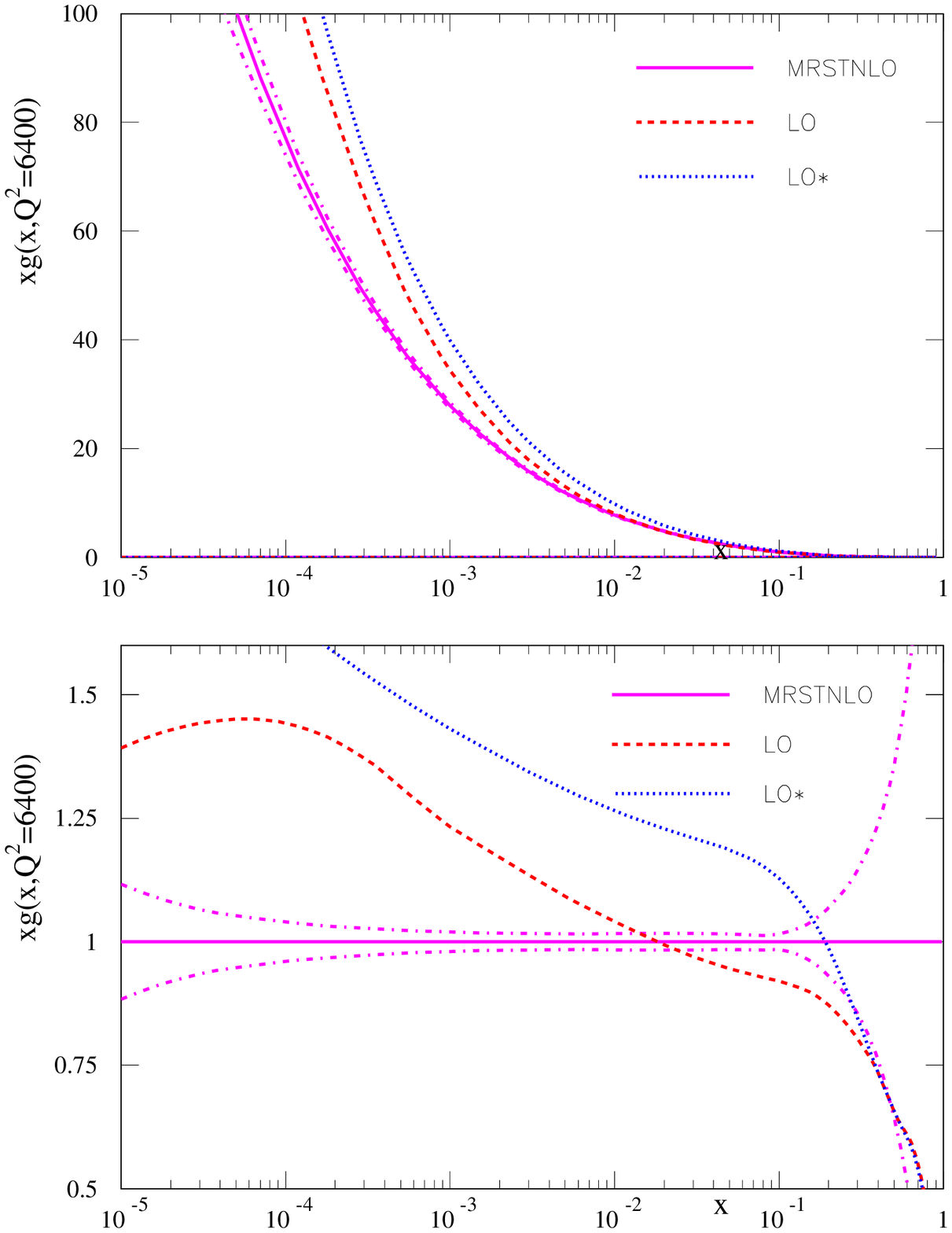}}
}
\vspace{0.3cm}
\caption{ 
Comparison between the LO, LO* and NLO up quarks (left) and between LO, LO* 
and NLO gluon (right). 
}
\label{lomompart}
\end{center}
\end{figure}

Relaxing the momentum sum rule at input\footnote{The evolution automatically 
conserves momentum.} and using the NLO definition of the strong coupling does 
dramatically improve the quality of the LO global fit (though a K-factor of 
1.3 is necessary for fixed target Drell-Yan data). The $\chi^2=3066/2235$ 
for the standard LO fit, and becomes $\chi^2=2691/2235$ for the modified fit. 
The data set fit is much the same as in \cite{Martin:2004ir} and the heavy flavour 
prescription used is that in \cite{nnlovfns}, which is a little different 
from that for previous fits at LO. There is a particularly big improvement in 
the quality of the fit to HERA data. When using the NLO definition of 
the coupling we obtain $\alpha_S(M_Z^2)=0.121$, which  is a much more sensible 
result than in the pure LO fit. The momentum carried by input partons goes up to $113\%$. We denote the 
partons resulting from this fit as the LO* parton distribution functions. 

The LO*, LO and NLO partons are compared in Fig.~\ref{lomompart}. One can 
see that the LO* quark distribution is bigger than at LO at $x < 0.1$, and 
only smaller than NLO for a small region centred slightly below $x=0.1$. 
Similarly $g(x,Q^2)$ is significantly bigger at LO* than at LO, and much 
bigger than NLO at small $x$.\footnote{Both LO and LO* gluon distributions 
are smaller than NLO at very large $x$. This is because the increase in the 
LO quark distributions in this region allows a good fit to Tevatron jet data 
without a significant high-$x$ gluon distribution. The LO gluon distribution 
at very high $x$ would be more similar to that at NLO in the DIS scheme 
\cite{Martin:2004ir}.} Hence, the LO* gluon distribution should do better than LO 
for gluon-gluon initiated processes (e.g. Higgs production) where $K$-factors 
are often much greater than unity. 

\begin{figure}
\begin{center}
\centerline{
\epsfxsize=0.7\textwidth\epsfbox{{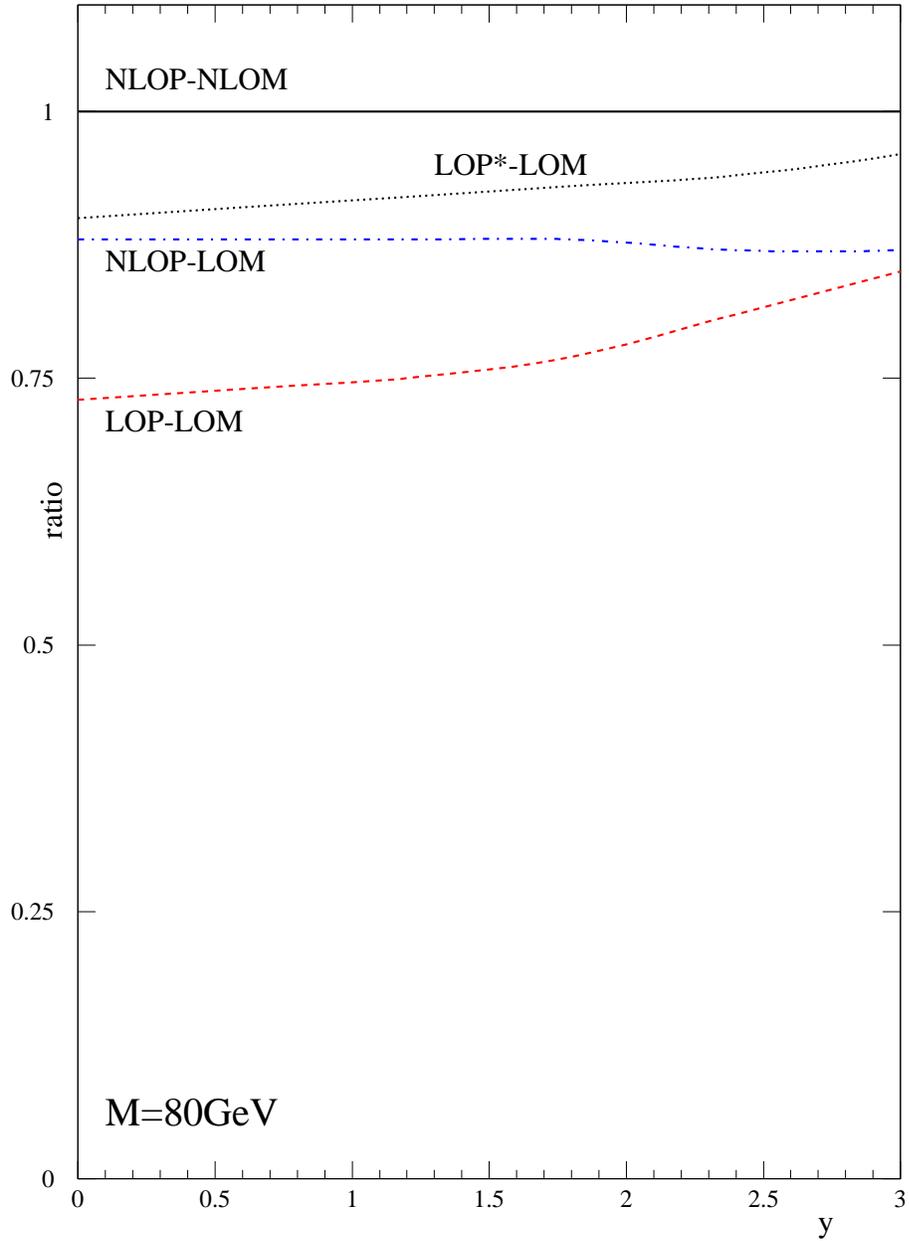}}
}
\vspace{0.3cm}
\caption{
Comparison of boson production at the LHC using combinations of different 
orders of parton distributions and hard partonic cross-sections, now including 
LO* pdfs. 
}
\label{monteraplhco}
\end{center}
\end{figure}

We can make a simple test of the potential of these improved LO* partons by 
repeating the comparisons of the previous section. Consider Drell-Yan 
production at the LHC, shown in Fig.~\ref{monteraplhco}. Indeed, we 
are nearer to the {\it truth} with the LO matrix element and LO* pdfs than 
with either LO or NLO pdfs. Moreover, the shape using the LO* pdfs is of 
similar quality to that using the NLO partons with the LO matrix element. 
So in this case LO* pdfs and NLO pdfs are comparably successful. 

\begin{figure}
\begin{center}
\centerline{
\epsfxsize=0.7\textwidth\epsfbox{{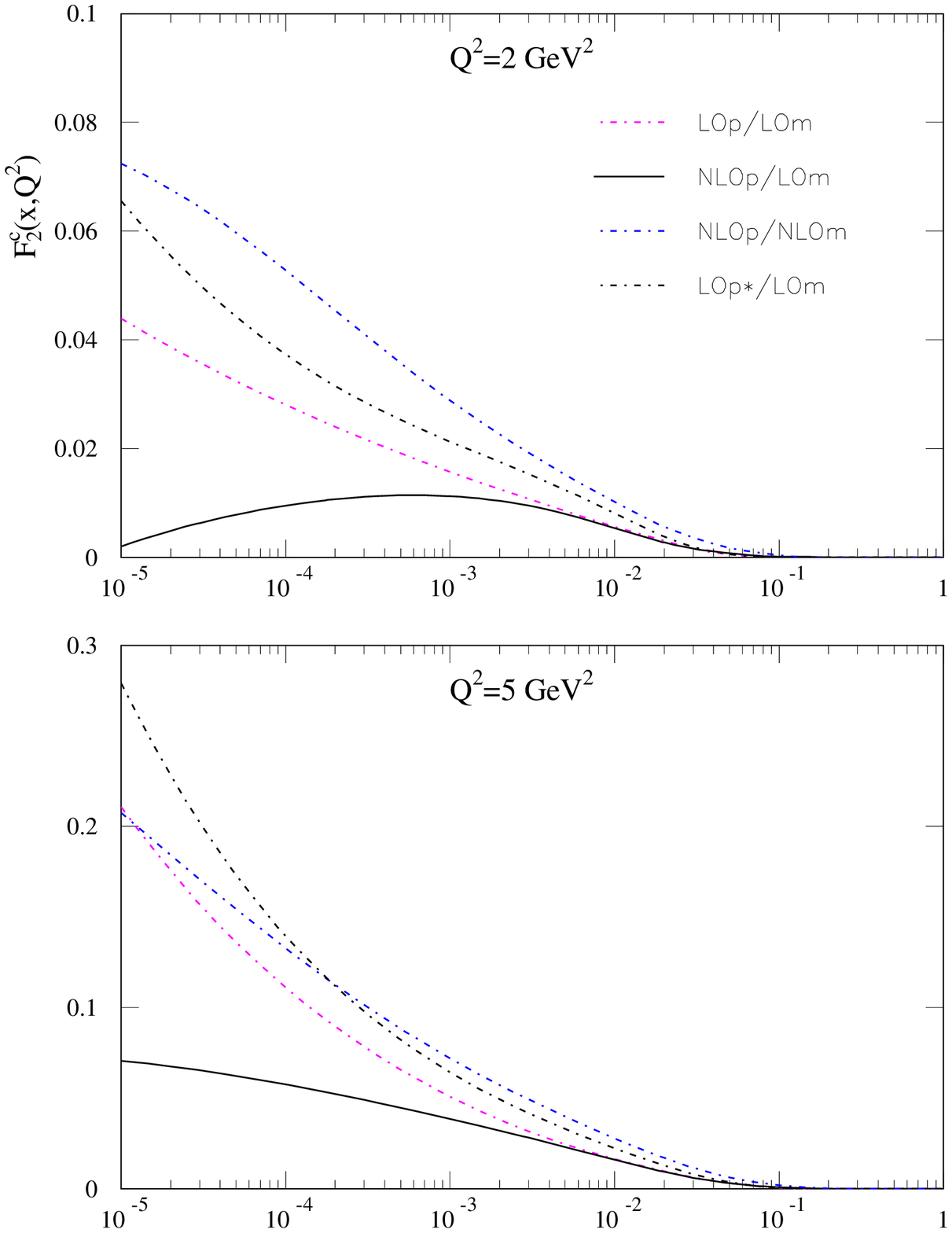}}
}
\vspace{0.3cm}
\caption{ 
A comparison of charm production at HERA using LO, NLO, and LO* parton 
distributions with LO coefficient functions and the {\it truth} of NLO pdfs 
with NLO coefficient functions. 
}
\label{charmlonloc}
\end{center}
\end{figure}

The exercise is also repeated for the charm structure function at HERA, as 
seen in Fig.~\ref{charmlonloc}. When 
using the LO coefficient function the LO* pdfs result is indeed nearest to 
the {\it truth} at low scales, being generally a slight improvement on the 
result using LO pdfs, and clearly much better than that using NLO pdfs. One 
would expect there to be a similar result for all processes dependent on the 
gluon distribution at small $x$ , e.g. hadro-production of $c$ and/or $b$ 
quarks at the LHC. 

These simple examples suggest that the LO* pdfs may well be a valuable tool 
for use with Monte Carlo generators at LO, combining much of the advantage 
of using the NLO distributions while avoiding the major pitfalls. However, 
the examples so far are rather unsophisticated. They are sufficient to show 
that something can go badly wrong in some cases, but in order to determine 
the best set of pdfs to use it is necessary to work a little harder. We need 
to examine a wide variety of contributing parton distributions, both in type 
of distribution and range of $x$. Also, the above examples are both 
completely inclusive, they have not taken into account cuts on the data. Nor 
have they taken account of any of the possible effects of parton showering, 
which is, of course, one of the most important features of Monte Carlo 
generators. Hence, before drawing any firm conclusions we will make a wide 
variety of comparisons for different types of possible process at the LHC, 
using Monte Carlo generators to produce the details of the final state.

\section{Details of Different Production Mechanisms} 
We consider a variety of final states for proton-proton collisions at LHC 
energies. In each case we compare the cross-section with LO matrix elements 
and full parton showering for the three cases of LO, LO* and NLO parton 
distributions. We also include the results using a Monte Carlo generator 
with NLO matrix element corrections \cite{Frixione:2002ik} together with 
NLO parton distributions, which we take to be the {\it truth}. If a process 
is not available in MC@NLO, we use the standard NLO approximation without 
the parton shower influence. The different examples span a range of values 
of $x$ and hard scales and probe different combinations of parton 
distributions. 


\begin{table}
  \centerline{\begin{tabular}{|c|c|c|c|c|}
  \hline
  Process                     &   LO generator &  NLO generator &  QCD scale  \\
                              &                &                &             \\
  \hline
    $pp \to W \to \mu\nu$     & CompHEP/HERWIG &    MC@NLO      &   $\hat s$  \\
    $pp \to Wj \to \mu\nu j$  & CompHEP/HERWIG &    MCFM        &   $\hat s$  \\
    $pp\to Z/\gamma\to2\mu$   & CompHEP/HERWIG &    MC@NLO      &   $\hat s$  \\
    $pp \to t,q$              & CompHEP/HERWIG &    MC@NLO      &   $m_t$     \\
    $g,g\to H$                & CompHEP/HERWIG &    MC@NLO      &   $m_H$     \\
    $pp\to Hqq$               & CompHEP/HERWIG &    VBFNLO      &   $m_H$     \\
    $pp\to \bb$               & HERWIG         &    MC@NLO      &   $\sqrt{m_b^2+<p_T>^2}$  \\
    $pp\to \tt$               & CompHEP        &    MC@NLO      &   $\sqrt{m_t^2+<p_T>^2}$  \\
    $pp\to jj$                & CompHEP/HERWIG &    JETRAD      &   $E_T$  \\
  \hline
\end{tabular}}
\caption{
A list of investigated processes at the LHC, along with 
Monte-Carlo codes and the hard QCD scales used in the calculations. 
}
\label{Tab:generators}
\end{table}

Table~\ref{Tab:generators} reports generators 
(\cite{Frixione:2002ik}, \cite{Boos:2004kh}, \cite{Belyaev:2000wn}, \cite{Corcella:2000bw}, 
\cite{Giele:1993dj}, and \cite{Figy:2003nv}) used and QCD scales applied in the calculations. 
As the main LO generator we use CompHEP interfaced with FORTRAN HERWIG, except $pp\to \bb$, 
where HERWIG is more appropriate generator. The main NLO generator is MC@NLO except three 
processes $pp\to Hqq$, $pp \to Wj \to \mu\nu j$ and $pp\to jj$, which are not available in 
MC@NLO. The following parameters are used in the calculations: 
$m_W = 80.4$ GeV, 
$m_Z = 91.176$ GeV, 
$\Gamma_W = 2.028$ GeV, 
$\Gamma_Z = 2.44$ GeV, 
$\cos {\theta_W} = 0.2311$, 
$m_b = 4.85$ GeV,
$m_c = 1.65$ GeV, 
$m_{\mu} = 105.66$ MeV, 
$\alpha = 1/127.9$ (fixed), 
running $\alpha_S$ (according chosen pdfs). 

\subsection{$W$ Production at the LHC.} 

\begin{table}
  \centerline{\begin{tabular}{|c|c|c|c|}
  \hline
  parton &   matrix   &     $\sigma$ (nb) & K-factor \\
         &  element   &                   &          \\
  \hline
    NLO  &     NLO    &      21.1        &          \\
    LO   &     LO     &      17.5        &   1.21   \\
    NLO  &     LO     &      18.6        &   1.13   \\
    LO*  &     LO     &      20.6        &   1.02   \\
  \hline
\end{tabular}}
\caption{
The total cross-sections $\sigma(pp\to W\to \mu\nu_{\mu})$ at the LHC. 
}
\label{Tab:Wcs}
\end{table}


\begin{table}
  \centerline{\begin{tabular}{|c|c|c|c|}
  \hline
  parton &   matrix   &     $\sigma$ (nb) & K-factor \\
         &  element   &                   &          \\
  \hline
    NLO  &     NLO    &      5.96         &          \\
    LO   &     LO     &      4.66         &   1.28   \\
    NLO  &     LO     &      4.20         &   1.42   \\
    LO*  &     LO     &      5.43         &   1.10   \\
  \hline
\end{tabular}}
\caption{
The total cross-sections $\sigma(pp\to Wj\to \mu\nu_{\mu}j)$ at the LHC. 
Cuts: $p_T(j) > 20$ GeV, $|\eta(j)| <$ 5.
}
\label{Tab:Wcsj}
\end{table}

We first revisit our previous LHC example of $W$ production, but with full 
parton showering and an investigation of the total cross-section and $p_T$ 
distribution as well as the pseudo-rapidity distribution. Hence, we consider 
the LO production mechanism of quark-antiquark annihilation where $x=0.006$ 
at central pseudo-rapidity. 
The total cross-sections are shown in Table~\ref{Tab:Wcs}, and we see that 
all predictions using the LO matrix elements are lower than the {\it truth}, 
but that using the LO* pdfs is easily closest. 

\begin{figure}
\begin{center}
\centerline{
\epsfxsize=17.0cm
\epsfysize=17.0cm
\epsffile{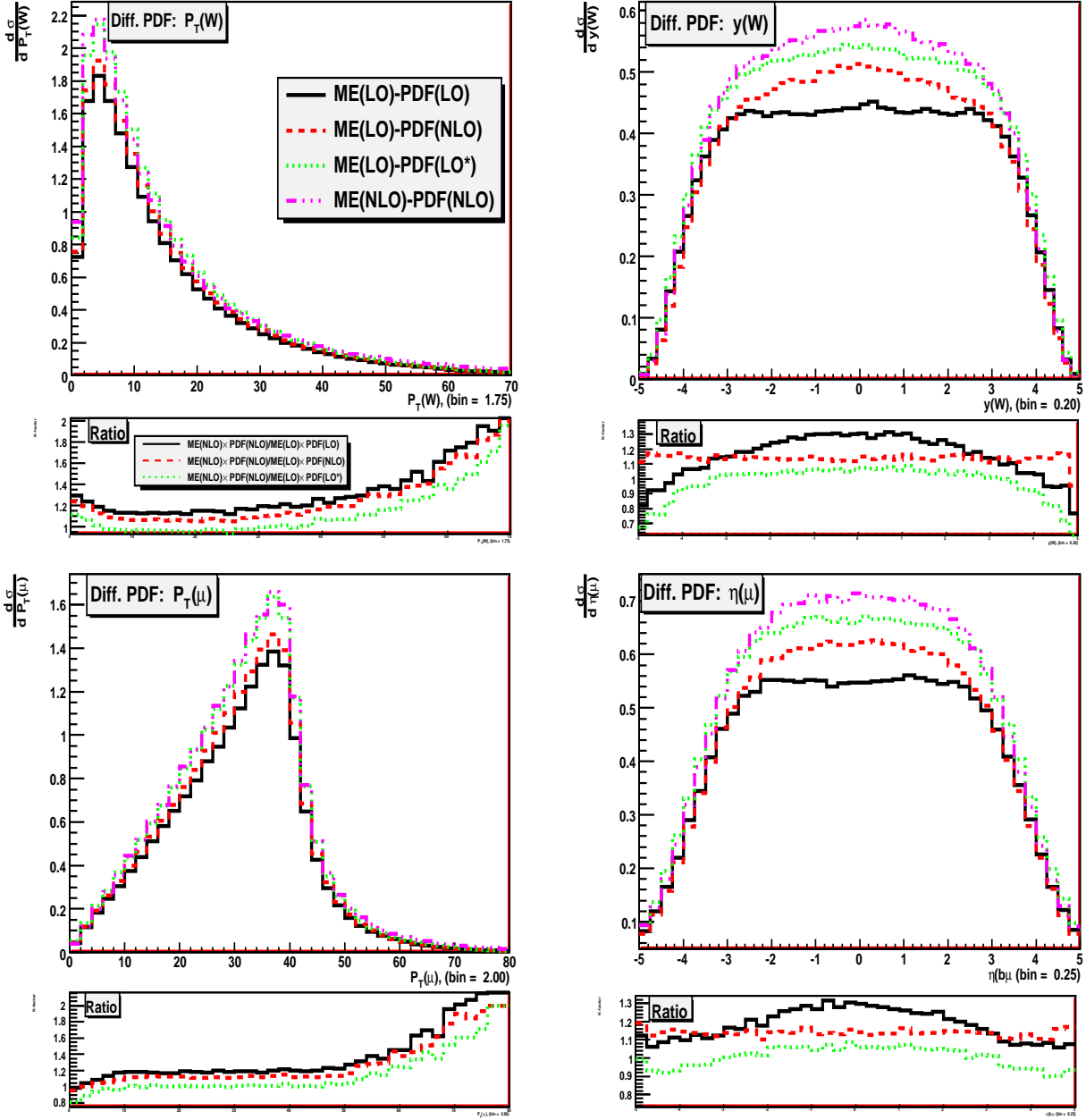}
}
\vspace{0.3cm}
\caption{ 
The comparison between the competing predictions for the differential 
\hspace{0.5cm}
cross-section for $W$-boson production at the LHC (left) and for the 
resulting muon (right). 
}
\label{Pic:w_ratio}
\end{center}
\end{figure}

The distributions are seen in the upper of Fig.~\ref{Pic:w_ratio}, where 
the small plots are the ratio of the {\it truth} to each of our LO matrix element 
predictions. For the W-boson pseudo-rapidity distribution the LO partons 
give a result which is suppressed at central rapidities due to the depletion of 
quarks seen before. The LO* partons 
give a good representation of the {\it truth}, being just a little small at central 
rapidity and a bit large at the very highest rapidity. The NLO partons are 
slightly worse in normalization, but overall are actually best in shape. They 
are a little low for very high rapidity, where the excess of LO* partons at very 
high and low $x$ overcompensates for enhancement in the NLO matrix elements 
while the NLO partons do not compensate at all. 
These high rapidities are outside the range of 
ATLAS and CMS, but may be relevant at LHCb. The LO partons give too low a 
result at central rapidity due to the depletion for $x \sim 0.01$. 

For the $p_T$ distributions none of the pdfs gives a good representation of 
the shape with $p_T$. In particular, they all lead to predictions which are 
too small with increasing large $p_T$. At LO the $p_T$ is simulated by a 
parton shower algorithm (in our case, by the HERWIG algorithm). By  
definition it underestimates the high-$p_T$ region in comparison with the true 
NLO correction. As we will see, this type of effect is a recurring feature. 
It cannot be solved by defining pdfs and must be borne in mind whenever using 
LO Monte Carlo generators. Nevertheless, all pdfs produce rather similar 
shapes using the LO generator, but again the LO* pdfs give the best 
normalization. 

In the lower of Fig.~\ref{Pic:w_ratio} we see the distributions for the 
final-state muon in the $W$ decay. The conclusions are similar. Again for the 
rapidity distribution the LO* distributions are best in normalization,  
and indeed are an excellent approximation to the {\it truth}. 
The rapidity of the muon is usually a bit higher than that of the parent $W$ 
so the effect right at the edge of the rapidity plot in the $W$ case is 
not visible for the muon, and would be smeared at even higher muon rapidity. 
The NLO partons are low in normalization, but arguably best in shape. Again the 
LO pdfs fail worst at central rapidity. For the $p_T$ distribution the 
NLO excess only sets in beyond $\sim 50 \GeV$, but this is simply due to the 
intrinsic $p_T$ of the muon from the $W$ decay. 

\begin{figure}
\begin{center}
\centerline{
\epsfxsize=1.0\textwidth\epsfbox{{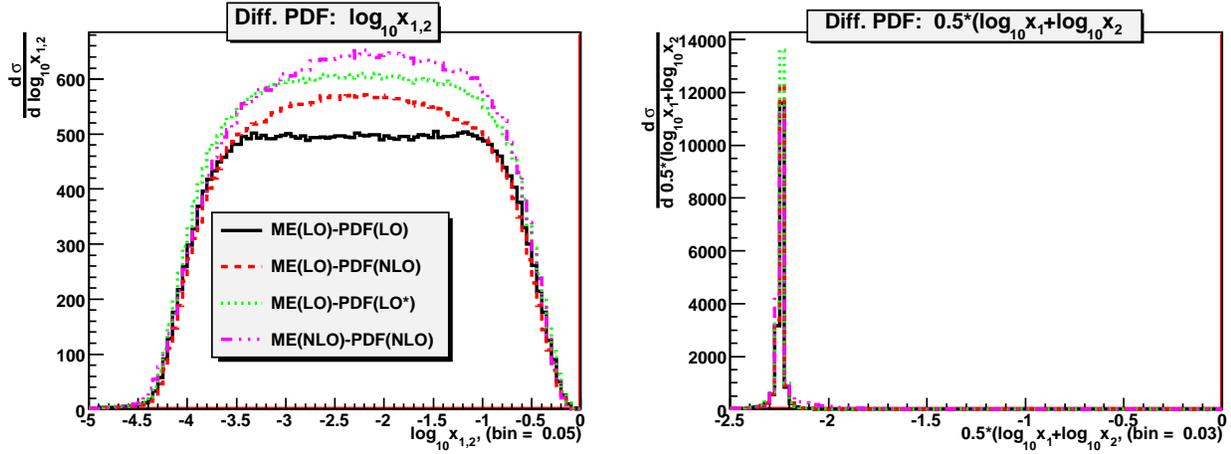}}
}
\vspace{0.3cm}
\caption{
The ranges in $x_1$ and $x_2$ of the contributing parton distributions 
for $W$-boson production at the LHC in the different types of calculation. 
}
\label{Pic:w_x}
\end{center}
\end{figure}

The range of $x$ and ``average'' $\log  \hat x = 0.5(log_{10}x_1+ log_{10}x_2)$ 
of the contributing partons are shown in Fig.~\ref{Pic:w_x}. For the 
LO production 
mechanism $\hat x$ is centred on the line $x_1x_2=m^2_W/s$ with some width, 
whereas at NLO there is a slightly increased 
contribution 
from the region $x_1x_2>m_W/\sqrt{s}$, where one of the incoming partons emits 
a relatively hard parton before undergoing the annihilation, this hard parton 
providing some hard $p_t$ to balance the $W$ boson. A depletion is seen for 
the LO partons in the region $x=0.001-0,1$. 

Since the LHC initial state is dominated by gluons, processes with gluons in 
the initial state are of particular interest. In the W-production process both 
initial partons are quarks. Initial gluons appear in the associated production 
of $W$ and one jet, and  in this process we probe both quark and gluon pdfs. We 
calculated the total cross-section for all our previously considered cases, 
the LO matrix element with three pdfs and the {\it truth}. The NLO approximation 
was calculated by the MCFM program~\cite{Campbell:2002tg}. Reasonable cuts were 
applied in order to exclude kinematic regions where the LO matrix element has 
singularities. The total cross-sections are reported in Table.~\ref{Tab:Wcsj}. As 
we can see, the LO* pdfs give the best normalization. 

\subsection{$Z/\gamma$ Production at the LHC.} 


\begin{table}
  \centerline{\begin{tabular}{|c|c|c|c|}
  \hline
  pdf type &   matrix   &     $\sigma$ (nb) & K-factor \\
           &  element   &                   &          \\
  \hline
    NLO  &     NLO    &      2.40          &             \\
    LO   &     LO     &      1.85          &   1.30      \\
    NLO  &     LO     &      1.98          &   1.26      \\
    LO*  &     LO     &      2.19          &   1.09      \\
  \hline
  \end{tabular}
  }
\caption{
The total cross-sections $\sigma(pp\to Z/\gamma\to \mu\mu)$ at the LHC with 
cuts ($p_T(\mu)>10\GeV$, $|\eta{\mu}|<5.0$). 
}
\label{Tab:Zcs}
\end{table}

Next we consider the rather similar process of quark-antiquark annihilation to 
$Z/\gamma$ bosons, decaying to muons. In order to exclude the dangerous region 
$m_{inv}(\mu\mu)\to 0$, where the matrix element at LO has a singularity, we 
apply some experimentally 
reasonable cuts cuts $p_T > 10\GeV$ and $|\eta|<5.0$. 
These cuts are more or less appropriate for most analyses in CMS/ATLAS. 
The process is dominated by the $Z$ peak. The mechanism is rather similar to 
that for $W$ production, but now the quark and antiquark are the same flavour 
and the $x$ at zero rapidity is slightly higher, i.e. $x_0=0.0065$. 

\begin{figure}
\begin{center}
\centerline{
\epsfxsize=17.0cm
\epsfysize=17.0cm
\epsffile{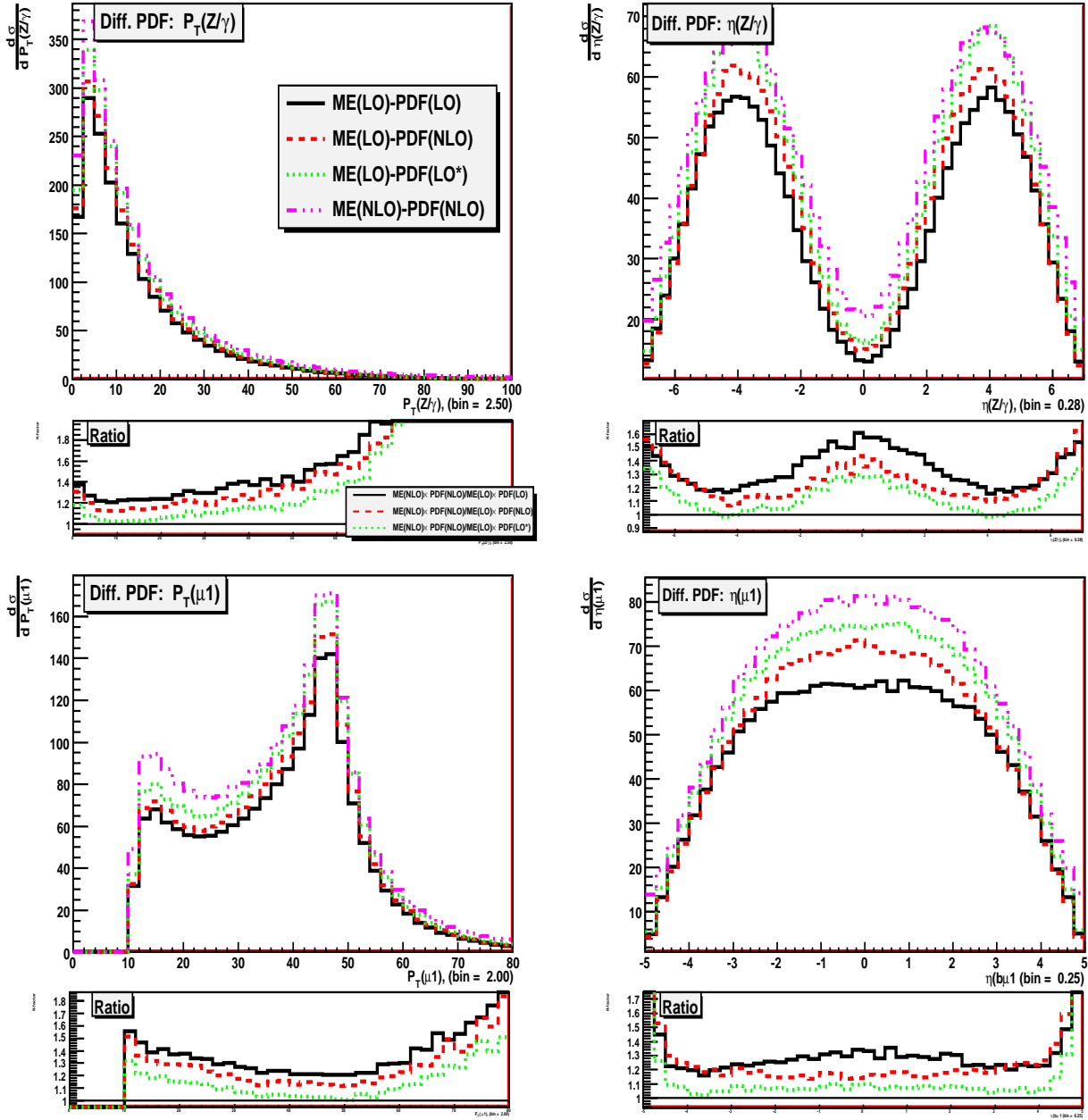}
}
\vspace{0.3cm}
\caption{ 
The comparison between the competing predictions for the differential 
cross-section for $Z/\gamma$-boson production at the LHC (upper plots) and for 
the resulting highest $p_t$ muon (lower plots). 
}
\label{Pic:z_ratio}
\end{center}
\end{figure}

The similarity is confirmed by the results. Again all the total cross-sections 
using the LO generators are lower than the {\it truth}, as seen in 
Table~\ref{Tab:Zcs}, but that using the LO* partons is easily closest. 
A similar conclusion holds for the distributions in terms of the final state boson or 
the highest-$p_T$ muon shown in the upper and lower plots of Fig.~\ref{Pic:z_ratio} 
respectively. For the boson the LO* partons gives comparable, perhaps marginally 
better, quality of shapes as the NLO partons, but better normalization. 
The LO partons have the worst suppression at central rapidity, and all partons 
give an underestimate of the high-$p_T$ tail. Similarly, for the muon the 
LO* partons give an excellent result for the rapidity distribution until 
$|\eta|>4$, better in shape and normalization that the NLO partons whilst 
the LO partons struggle at central $\eta$. Again, as in $W$ production, the 
$p_T$ distribution of the muon is better than for the boson, and in normalization 
is best described by the LO* pdfs. 

\begin{figure}
\begin{center}
\centerline{
\epsfxsize=1.0\textwidth\epsfbox{{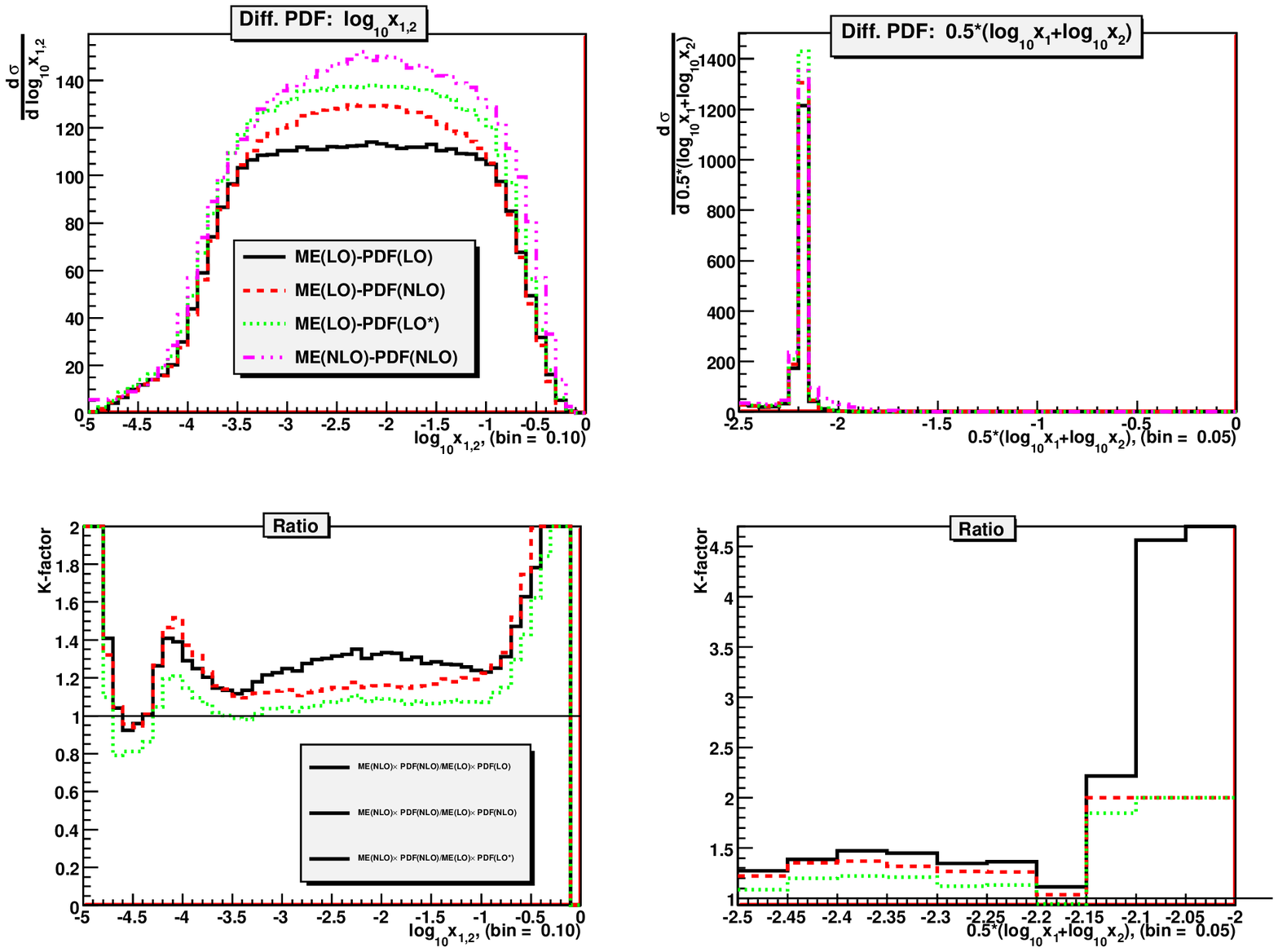}}
}
\vspace{0.3cm}
\caption{ 
The distributions of $x_{1,2}$ of the contributing parton distributions 
for $Z/\gamma$-boson production at the LHC in the different types of 
calculation. 
}
\label{Plot:z_x}
\end{center}
\end{figure}
We illustrate the ranges of $x_1$ and $x_2$ in in Fig.~\ref{Plot:z_x}. The 
upper right-hand plot shows that the peak in the partons contributing is indeed at 
$x_1x_2 = m_Z^2/s$, with some contribution from the region of 
higher $x_1x_2$. Again, this is particularly for the case of the NLO matrix 
element, where one parton can make a fairly hard emission before the 
annihilation process. This contribution from higher $x_1x_2$ is most clearly 
seen in the 
lower right-hand plot. The upper left-hand plot shows the contributing $x$ of 
each parton and is dominated by the line $x_1x_2 \approx m_Z^2/s$, 
with the plateau falling off when the quark distribution for the higher 
$x$ quark falls away for $x \to 1$, i.e. the contribution is small when  
$x_1 \approx 0.3$ so $x_2 \approx 0.0001$. Perhaps most interesting is the 
lower left-hand plot which shows the ratio of the contribution from particular
values of $x$ for the {\it truth} divided by each of the LO generator 
calculations. In all cases the ratio diverges at very high $x$ due to the 
possibility of radiation of a hard parton before annihilation at NLO. However,
over the main region of contribution to the cross-section, i.e. $ 0.001<x<0.1$,
the K-factor for the NLO and LO* pdfs is flat and close to 1 (though closer 
for the LO* pdfs), whereas there is a distinct bump to values greater than 1
for the LO pdfs. This illustrates very clearly the depletion of the 
quarks in this region for the LO partons, and is very similar in the previous 
case of $W$ production. 

\subsection{Single Top Production at the LHC.} 

\begin{table}
  \centerline{\begin{tabular}{|c|c|c|c|}
  \hline
  pdf type &   matrix   &     $\sigma$ (pb) & K-factor \\
           &  element   &                   &          \\
  \hline
     NLO   &     NLO    &   259.4           &          \\
     LO    &     LO     &   238.1           &  1.09    \\
     NLO   &     LO     &   270.0           &  0.96    \\
     LO*   &     LO     &   297.5           &  0.87    \\
  \hline
\end{tabular}}
\caption{
The total cross-sections for the single top production (with top decay 
$t\to \mu\nu_{mu},b$) at the LHC. 
}
\label{Tab:STcs}
\end{table}

Now we consider a somewhat different process, i.e. single top production 
with the top decaying $t \to \mu^{+}+\nu_{\mu} + b$. At the partonic level 
the dominant interaction process is $qb\to qt$ (or $q\bar b\to q\bar t$), 
where the $b$-quark has been emitted from the gluon. Since the $b$-quark pdf 
is determined by evolution from the gluon pdfs, this cross-section probes 
both the gluon distribution and the quark distributions for invariant masses 
of above about $200\GeV$, i.e. at central rapidity $x_0 \sim 0.01-0.1$. 
The $t$-channel nature of this process makes the invariant mass of the 
final state and the correspondingly probed $x$ values less precise than the 
previous case. The total 
cross-section for the various methods of calculation is seen in 
Table~\ref{Tab:STcs} (we excluded Br($t \to \mu^{+}+\nu_{\mu} + b$)). 
In this case the result using the LO generators and the LO pdfs is 
suppressed, but that using the LO* pdfs is now larger than 
the {\it truth}. This is due to the large enhancement of the LO* gluon 
distribution. The NLO pdfs give the closest normalization. 

\begin{figure}
\begin{center}
\centerline{
\epsfxsize=17.0cm
\epsfysize=17.0cm
\epsffile{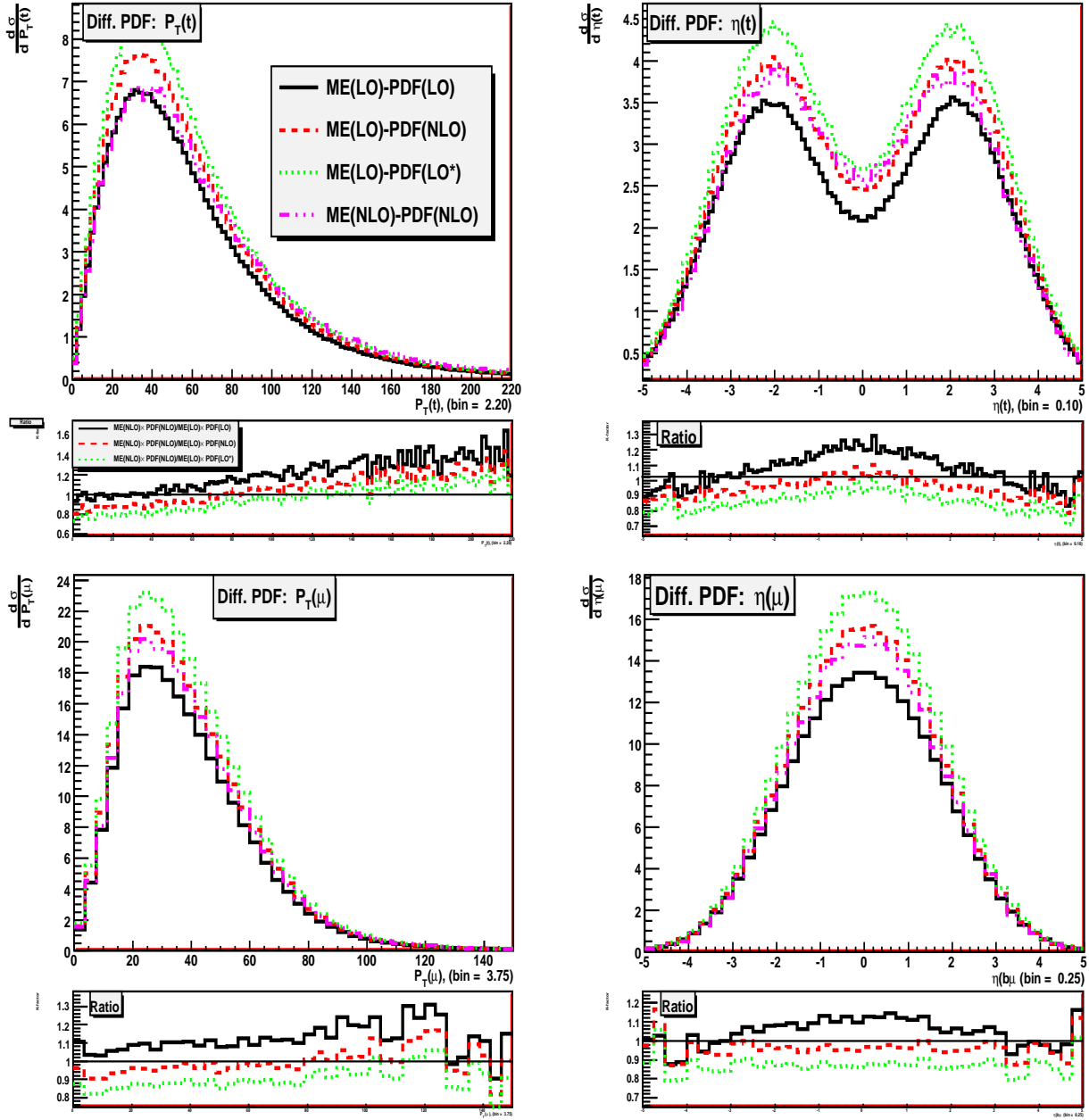}
}
\vspace{0.3cm}
\caption{ 
The comparison between the competing predictions for the differential 
cross-section for single top production at the LHC (upper plots) and for 
the resulting $p_t$ muon (lower plots). 
}
\label{Plot:st_phys}
\end{center}
\end{figure}

\begin{figure}
\begin{center}
\centerline{
\epsfxsize=1.0\textwidth\epsfbox{{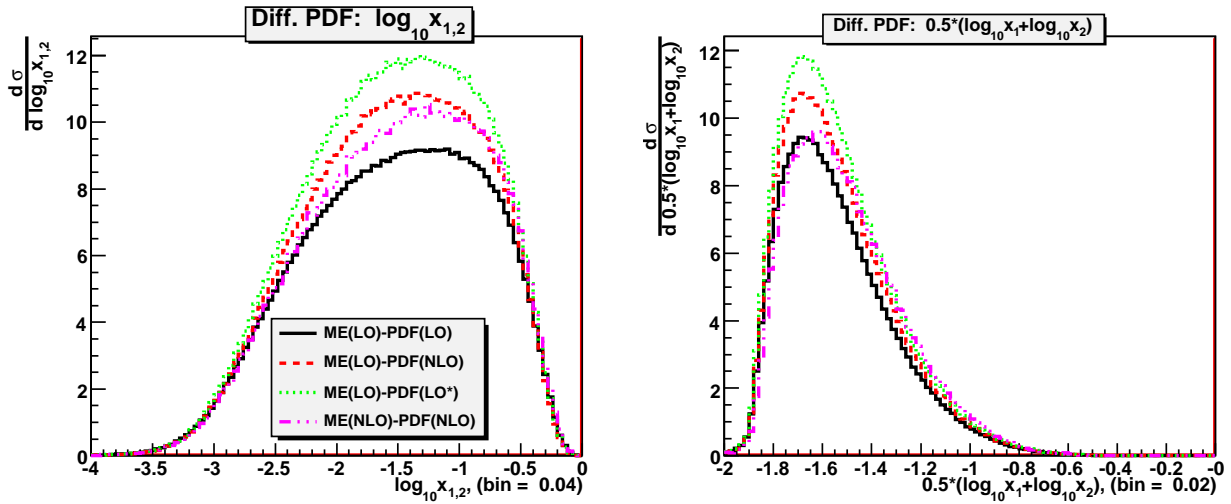}}
}
\vspace{0.3cm}
\caption{ 
The distributions of $x_{1,2}$  of the contributing parton distributions for 
the single top production at the LHC in the different types of calculation. 
}
\label{Plot:st_x}
\end{center}
\end{figure}

The distributions in terms of $p_T$ and $\eta$ of the final state top and 
$\mu$ originated from the top are shown in Fig.~\ref{Plot:st_phys}. 
For the top distribution the result using the LO generator and the LO* and 
NLO pdfs give a very similar result, being better than the LO pdf result both 
for normalization and for shape due to the suppression of the LO quarks at 
central rapidities. In the case of the $\mu$ the distributions calculated with 
the LO generator look better than for the top, since the real NLO correction 
(irradiation of an extra parton) plays a lesser role 
for the top decay products. 
In this process there is a particular NLO enhancement at central rapidity, so 
it gives a total cross-section larger than the {\it truth}. 
In Fig.~\ref{Plot:st_x}
we display the ranges of $x$ sampled. The right-hand plot 
shows the main difference in this process as compared 
with the previous cases. Here we have much larger values of ``average'' $x$. 

The process also illustrates the importance of the NLO corrections in experimental 
analyses. If we accept the picture where $qb\to qt$ is the LO approximation 
for 
the process\footnote{There exists another prescription, where one considers 
$qg\to qt+\bar b$ as a LO approximation~\cite{Stelzer:1997ns}. From our point 
of view, this approach is less motivated both from the mere vertex calculation 
(both Feynman graphs of the process have 3 vertices, whereas $qb\to qt$ has 
only two vertices) and availability of the $b$-quark pdf. The $b$ 
distribution takes 
into account some logarithmic corrections which cannot be implemented 
in the LO matrix element.}, 
the main correction comes from the gluon splitting to $\bb$, where one 
$b$-quark participates in the top production and the second $b$ is a 
parton-spectator. 
Thus, at LO we have the $b$-quark in the initial state, but we assume no 
intrinsic 
$b$-quarks in protons (as we pointed out the $b$ 
pdfs are calculated based on the 
gluon distribution). It means we must produce the second $b$-quark somehow in 
any reliable simulation. Usually Monte-Carlo programs generate the parton via 
the initial parton shower approach. It is clear the matrix element calculation 
gives a more central $b$-quark with higher $p_T$. So, if the parton is 
important 
in an analysis, the LO approximation does not work well whatever pdf
we apply\footnote{In fact, one can combine the LO approximation and the 
main real correction due to $g\to \bb$. See more details in~\cite{Boos:2006af}.}. 

\subsection{Higgs Production at the LHC -- Gluon-Gluon Fusion.}

\begin{table}
  \centerline{
  \begin{tabular}{|c|c|c|c|}
  \hline
  pdf type &   matrix   &     $\sigma$ (pb) & K-factor \\
           &  element   &                   &          \\
  \hline
     NLO   &     NLO    &      38.0         &          \\
     LO    &     LO     &      22.4         &  1.70    \\
     NLO   &     LO     &      20.3         &  1.87    \\
     LO*   &     LO     &      32.4         &  1.17    \\
  \hline
  \end{tabular}
  }
\caption{
The total cross-sections $\sigma(pp\to H)$ at the LHC. Strictly speaking 
this is $pp\to H\to\tau\bar\tau$ with BR($H\to\tau\bar\tau$) excluded. 
}
\label{Tab:H_bb_cs}
\end{table}

\begin{figure}
\begin{center}
\centerline{
\epsfxsize=17.0cm
\epsfysize=17.0cm
\epsffile{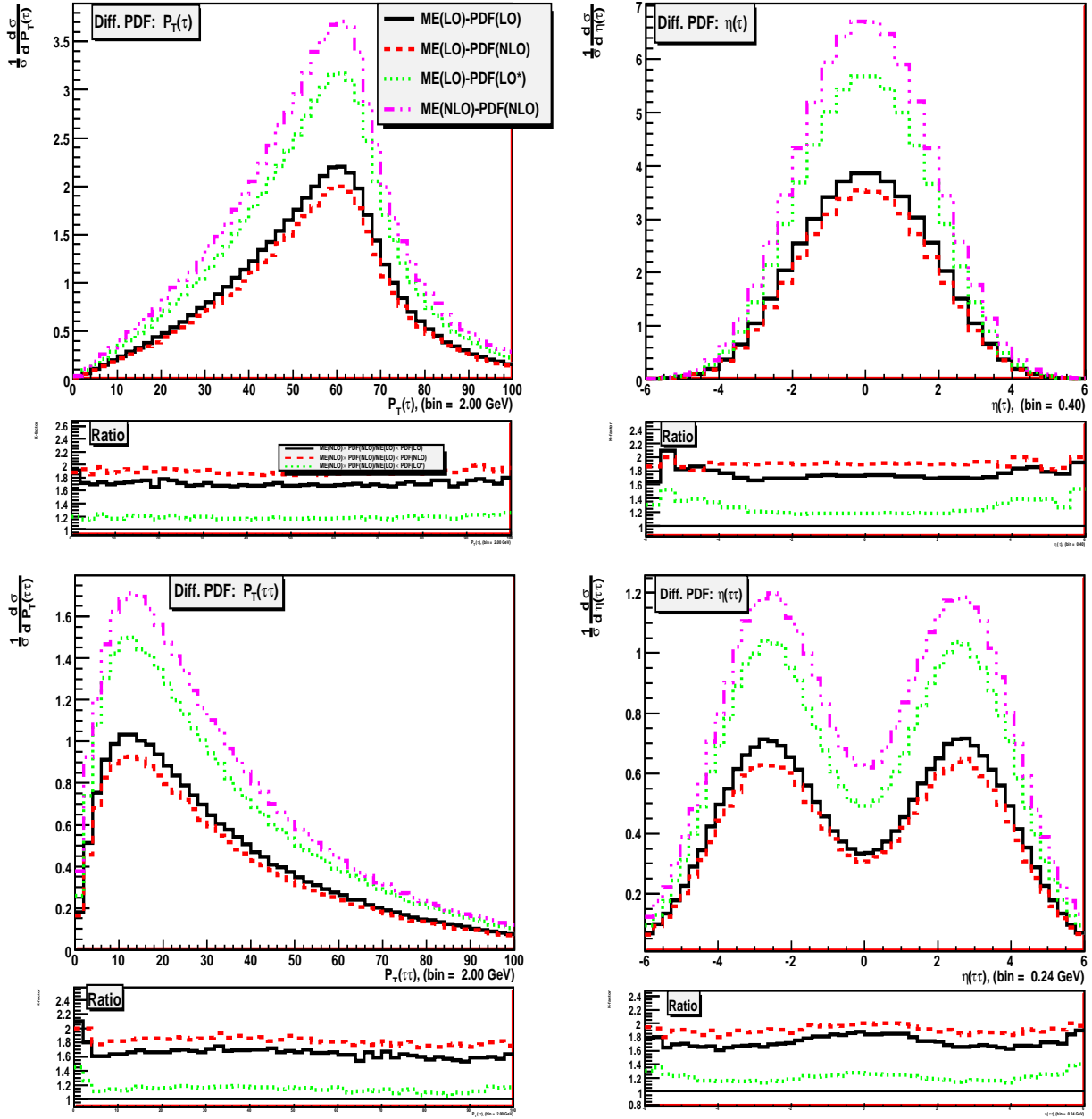}
}
\vspace{0.3cm}
\caption{ 
The comparison between the competing predictions for the differential 
cross-section for the Higgs production at the LHC (lower plots) and for 
the resulting $\tau$ lepton (upper plots). 
}
\label{Plot:h_phys}
\end{center}
\end{figure}

\begin{figure}
\begin{center}
\centerline{
\epsfxsize=1.0
\textwidth\epsfbox{{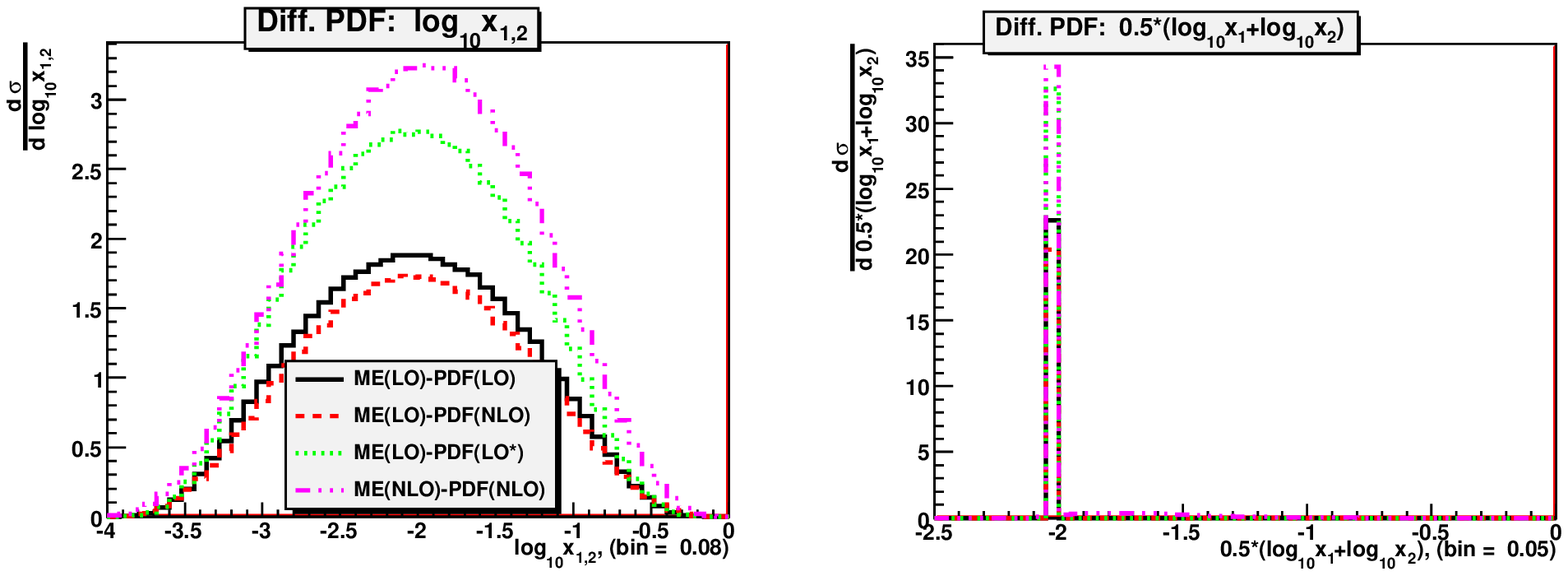}}
}
\vspace{0.3cm}
\caption{ 
The distributions of $x_{1,2}$  of the contributing parton distributions for 
the process $pp\to H\to\tau\bar\tau$ at the LHC in the different types of 
calculation. 
}
\label{Plot:H_x}
\end{center}
\end{figure}

We now consider a process which is determined entirely by the gluon 
distribution, 
i.e. production of Higgs of mass $130\GeV$ from gluon-gluon fusion. So the 
cross-section depends on the gluon distribution with the $x$ at central 
rapidity being $x_0=0.01$. It is well known that there is a very 
large K-factor, 
approximately 1.7, for Higgs production via this mechanism, so it is 
no surprise 
that when using the LO generator the cross-section is suppressed by 
roughly this 
factor using both the LO and NLO pdfs, whose gluon is of a similar size for 
of $x\sim 0.01$. However, from the right-hand side of Fig.~\ref{lomompart} 
we see that the LO* gluon distribution is enhanced by a factor of 
$\approx 1.25$ for the relevant value of $x$ and the extra gluon contribution 
factor of $1.25^2$ compensates a large part of the NLO K-factor. Hence, 
the result using the LO* pdfs is much better than for the LO and NLO pdfs,
as seen in Table~\ref{Tab:H_bb_cs}. 
Since $gg\to H$ is an $s$-channel process, the ``average'' $x$ has the same 
profile as the $W$ and $Z/\gamma$ production, as we see on the right-hand 
plot in Fig.~\ref{Plot:H_x}. The distributions in the final state are shown 
in Fig.~\ref{Plot:h_phys}. The shapes are good in all cases, perhaps best 
for NLO pdfs, but the normalization is poor except for the LO* pdfs. 

\subsection{Higgs Production at the LHC -- Vector Boson Fusion.}

\begin{table}
  \centerline{\begin{tabular}{|c|c|c|c|}
  \hline
  pdf type &   matrix   &     $\sigma$ (pb) & K-factor \\
           &  element   &                   &          \\
  \hline
     NLO   &     NLO    &      4.52         &          \\
     LO    &     LO     &      4.26         &  1.06    \\
     NLO   &     LO     &      4.65         &  0.97    \\
     LO*   &     LO     &      4.95         &  0.91    \\
  \hline
\end{tabular}}
\caption{
The total cross-sections for the process $pp\to Hqq$ at the LHC. 
}
\label{Tab:Hqq_cs}
\end{table}

\begin{figure}
\begin{center}
\centerline{
\epsfxsize=1.0\textwidth\epsfbox{{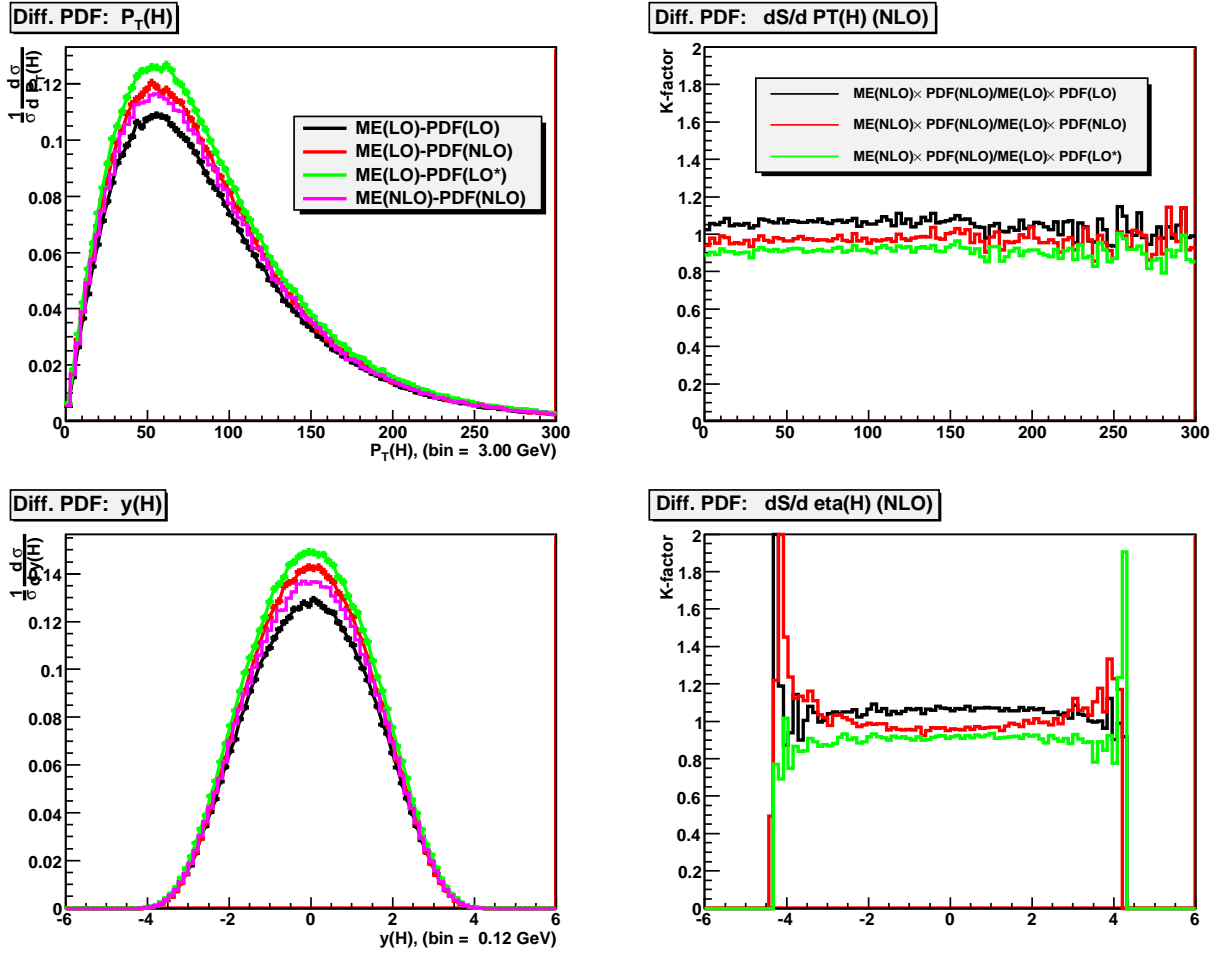}}
}
\vspace{0.3cm}
\caption{ 
The comparison between the competing predictions for the differential 
cross-section for the process $pp\to Hqq$ at the LHC. 
}
\label{Plot:Hqq_phys}
\end{center}
\end{figure}

\begin{figure}
\begin{center}
\centerline{
\epsfxsize=1.0
\textwidth\epsfbox{{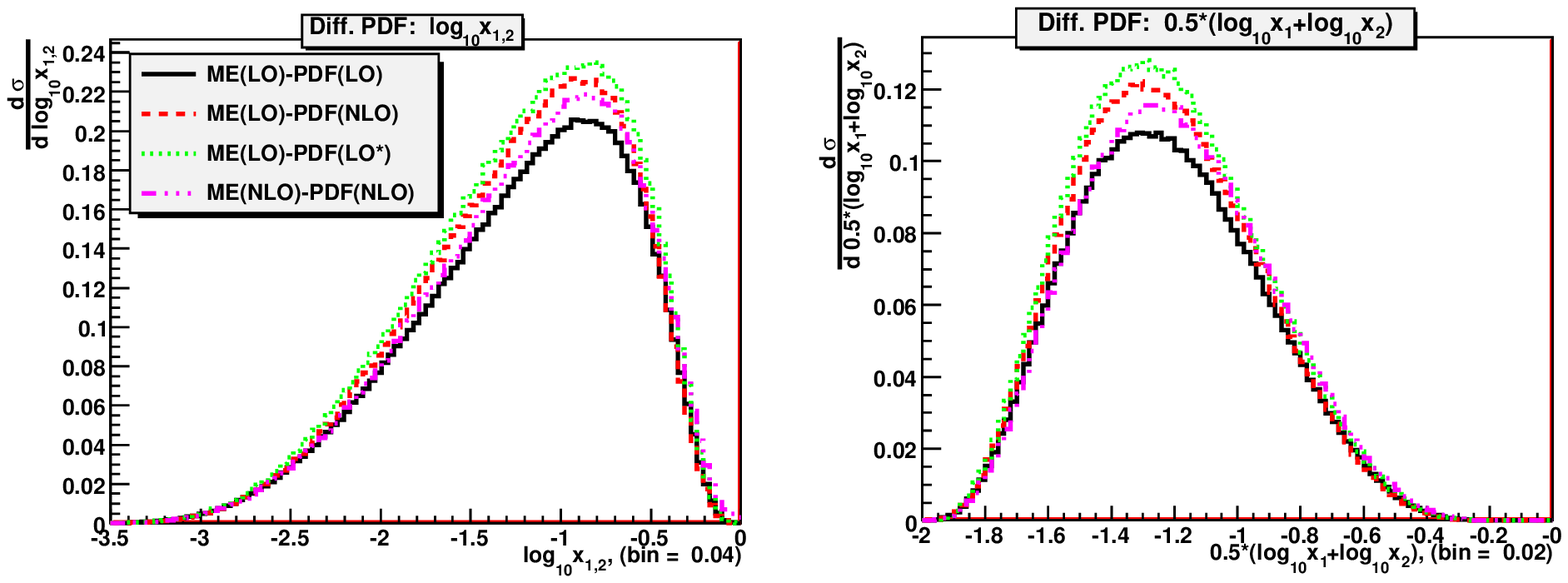}}
}
\vspace{0.3cm}
\caption{ 
The distributions of $x_{1,2}$  of 
the contributing parton distributions for the process $pp\to Hqq$ 
at the LHC in the different types of calculation. 
}
\label{Plot:Hqq_x}
\end{center}
\end{figure}

We again consider Higgs production, but via a different mechanism, i.e. a 
quark from each proton emits a vector boson which fuse to produce the final 
state Higgs boson. In the case we use quark pdfs and probe a different 
$x$ region $\approx 0.1$. As we can see from Table.~\ref{Tab:Hqq_cs}, in 
this case the NLO K-factor is only a few percent (positive or negative), 
so the result using the LO generator and the NLO pdfs is only slightly above 
the {\it truth}. The result using the LO
quarks is about $6\%$ too low due to the suppression of the quarks. We note 
that the LO* quarks are similar to the NLO quarks for $x=0.01$ so they give a 
very similar total cross-section, i.e. just a little above the {\it truth}. 
The rapidity distributions are very good using both the LO* and NLO partons as
seen in Fig.~\ref{Plot:Hqq_phys}
but we have, as usual, a 
small underestimate at central rapidities when using the 
LO pdfs. In this case the NLO corrections do little to the shape of 
the $p_T$ distribution. The values of $x$ probed are seen in 
Fig.~\ref{Plot:Hqq_x} and go to quite high values of $x$. 

\subsection{Heavy Quark Production at the LHC.}

\begin{table}
  \centerline{\begin{tabular}{|c|c|c|c|}
  \hline
  pdf type &   matrix   &     $\sigma$ ($\mu$b) & K-factor \\
           &  element   &                   &          \\
  \hline
     NLO   &     NLO    &      2.76         &          \\
     LO    &     LO     &      1.85         &  1.49    \\
     NLO   &     LO     &      1.56         &  1.77    \\
     LO*   &     LO     &      2.63         &  1.05    \\
  \hline
\end{tabular}}
\caption{
The total cross-sections $\sigma(pp\to \bb)$ at the LHC. 
Applied cuts: $p_T>20\GeV$, $|\eta(b)|<5.0$, and $\Delta R(b,\bar b)>0.5$. 
}
\label{Tab:cs_bb}
\end{table}

\begin{table}
  \centerline{\begin{tabular}{|c|c|c|c|}
  \hline
  pdf type &   matrix   &     $\sigma$ (pb) & K-factor \\
           &  element   &                   &          \\
  \hline
     NLO   &     NLO    &      812.8        &          \\
     LO    &     LO     &      561.4        &  1.45    \\
     NLO   &     LO     &      531.0        &  1.53    \\
     LO*   &     LO     &      699.4        &  1.16    \\
  \hline
\end{tabular}}
\caption{
The total cross-sections $\sigma(pp\to\tt)$ at the LHC. 
}
\label{Tab:cs_tt}
\end{table}

We now consider $\bb$ and $\tt$ production at the LHC. Although the dominant 
contribution at LO, $gg/q\bar q\to \bb$ does not have soft and collinear 
singularities (the process is called Flavour Creation or FCR), there are 
two other sources of $b$-quarks at hadron colliders at LO: $qb\to qb$, where 
the 
second $b$-quark is simulated by initial parton showers (the process is called 
Flavour Excitation, or FEX), and the QCD $2\to2$ process with light quarks 
and/or gluons, where the $b$-quark pair arises in the initial or final parton 
showers\footnote{For example, the total cross-section for the improved LO 
pdfs from Table~\ref{Tab:cs_bb} can be separated into three terms: 
$\sigma_{tot} =\sigma_{FCR} + \sigma_{FEX} + \sigma_{GSL}$, 
where $\sigma_{FCR} 
= 1.6\;\mu$b, $\sigma_{FEX} = 0.57\;\mu$b, and $\sigma_{GSP} = 0.46\;\mu$b -- 
the total cross-sections for the FCR, FEX, and GSP processes respectively.} 
(called Gluon Splitting or GSP). These latter processes have massless 
partons and, thus, soft and collinear singularities. In order to exclude 
the dangerous regions, where the LO approximation does not work, we apply 
some reasonable cuts: $p_T(b)>20\GeV$, $|\eta(b)|<5.0$, 
$\Delta R(b,\bar b)>0.5$. 
It is interesting that we cannot separate the subprocesses at NLO, so only 
the FCR process exists at NLO (see more details in~\cite{Frixione:2003ei}). 

\begin{figure}
\begin{center}
\centerline{
\epsfxsize=1.0
\textwidth\epsfbox{{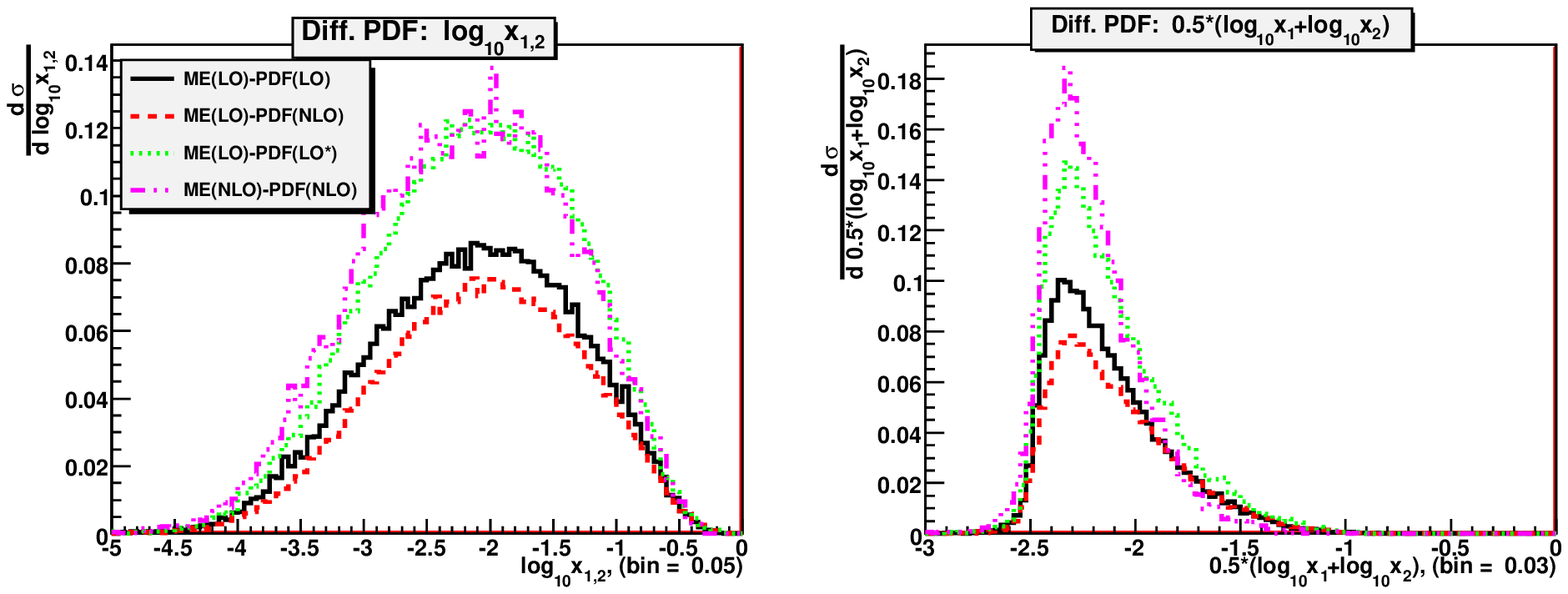}}
}
\vspace{0.3cm}
\caption{ 
The distributions of $x_{1,2}$ of the contributing parton distributions for 
the process $pp\to \bb$ at the LHC in the different types of calculation. 
}
\label{Plot:bb_x}
\end{center}
\end{figure}

At the LHC the dominant initial state in this case is gluon-gluon, similar 
to the inclusive Higgs, but at much lower $p_T$ (which dominates the total 
cross-section). So, in the process we probe rather 
low $x \sim 10^{-3}-10^{-2}$ (see Fig.~\ref{Plot:bb_x}). 
The total cross-sections are shown in Table~\ref{Tab:cs_bb}. In this case
we are at low enough $x$ to be sensitive to the small-$x$ divergence in the 
NLO matrix elements, and the NLO correction is very large. All of the results 
obtained using the LO generator are below the {\it truth}, but the reduced NLO
gluon means that the NLO pdfs give by far the worst result. The best absolute 
prediction is obtained using the LO* partons. In this case the LO gluon 
distribution is larger than at NLO, so LO pdfs give the second-best result.

\begin{figure}
\begin{center}
\centerline{
\epsfxsize=17.0cm
\epsfysize=17.0cm
\epsffile{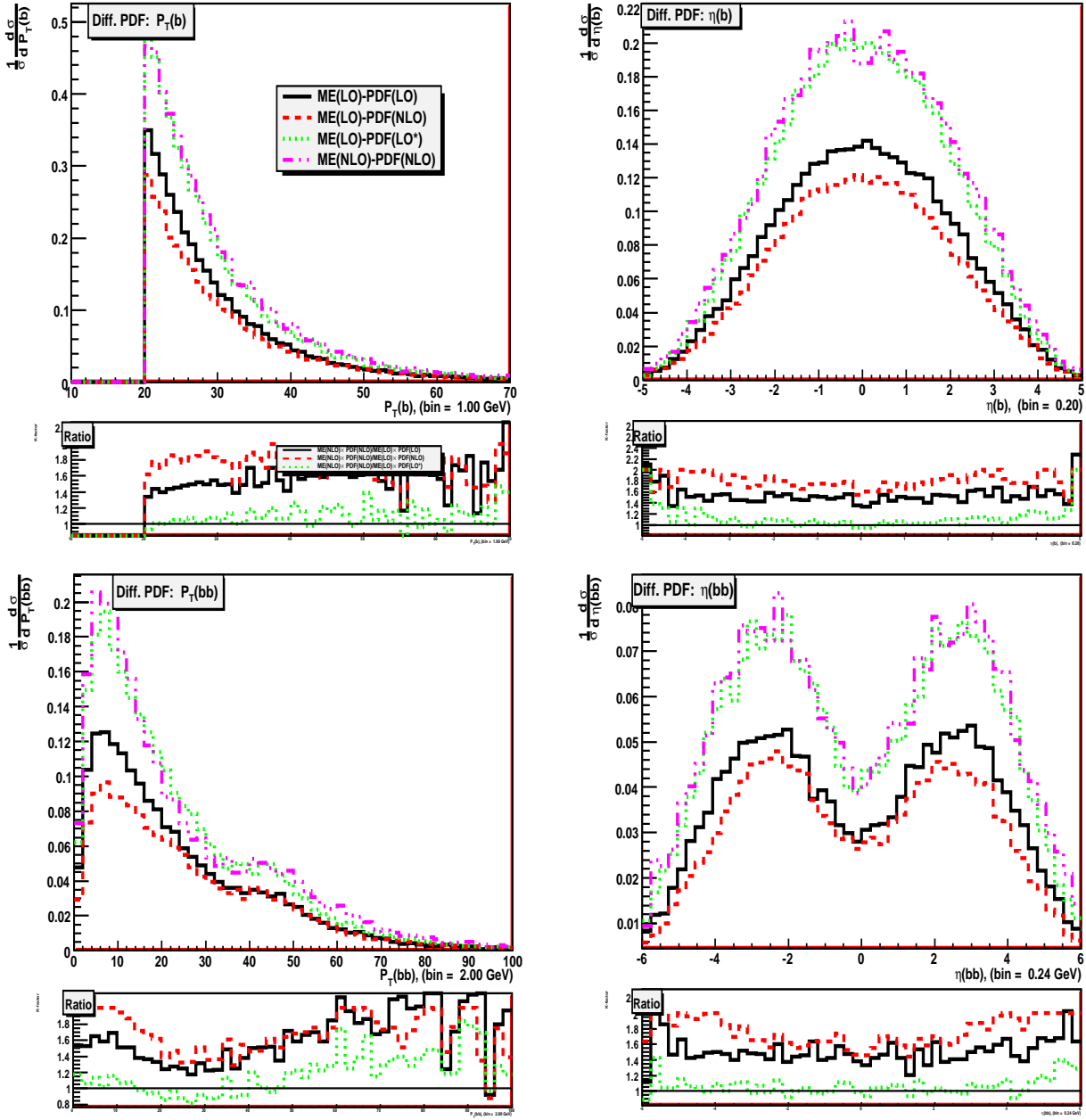}
}
\vspace{0.3cm}
\caption{ 
Differential cross-sections for $b$ production at the LHC (upper plots) 
and for a $\bb$ pair (lower plots).
}

\label{Plot:bb_phys}
\end{center}
\end{figure}

The differential distributions in terms of $p_T$ and $\eta$ of a single
$b$ quark are shown on the upper plots and for the pseudo-rapidity and $p_T$ 
of a $\bb$ pair on the lower plots in Fig.~\ref{Plot:bb_phys}. When 
using the LO generator the LO* pdfs do well for the single $b$ rapidity 
distribution, but underestimate a little at high rapidity. The LO and NLO 
pdfs are similar in shape, but the normalisation is worse. All pdfs obtain 
roughly the right shape for the $\eta(\bb)$, except small 
underestimation at very high rapidity. 
However, for all partons there is a small problem with the shape as a 
function of 
$p_T$. Obviously, all the ratio curves become higher as $p_T$ goes up. 
As for previous processes this happens due to the different behaviour of the 
additional parton generated at NLO compared to those 
generated by parton showers. 
In general, we conclude the LO* pdfs give the best results in the comparison. 
Unlike the inclusive Higgs this is not a pure $s$-channel process, 
and, as we can see 
in the right-hand plot in Fig.~\ref{Plot:bb_x}, there is no clear peak for 
``average'' $x$, but the main contribution in the process comes from the 
very low 
$x$ region, near $10^{-2.5}-10^{-2}$. In this region the LO* approach works 
better than in the high $x$ region. 

Another interesting heavy quark production process is double top quark 
production. As a short check we calculated the total cross-section for the 
process. Table~\ref{Tab:cs_tt} reports the numbers. At the LHC this 
process is dominated by the gluon contribution $gg\to t\bar t$. For example, 
$\sigma_{ME[LO]-PDF[LO]}=
\sigma_{gg\to t\bar t}+\sigma_{q\bar q\to t\bar t}=486.9\;pb+74.5\;pb$. 
The LO* pdfs appreciably enlarge the gluonic cross-section, namely, 
$\sigma_{ME[LO]-PDF[LO*]}=
\sigma_{gg\to t\bar t}+\sigma_{q\bar q\to t\bar t}=622.1\;pb+77.3\;pb$. Again the 
LO* pdfs give the best prediction.

\subsection{Jet Production at the LHC.}

\begin{table}
  \centerline{\begin{tabular}{|c|c|c|c|}
  \hline
  pdf type &   matrix   &     $\sigma$ ($\mu$b) & K-factor \\
           &  element   &                   &              \\
  \hline
     NLO   &     NLO    &     183.2         &              \\
     LO    &     LO     &     149.8         &      1.22     \\
     NLO   &     LO     &     115.7         &      1.58     \\
     LO*   &     LO     &     177.5         &      1.03     \\
  \hline
\end{tabular}}
\caption{
The total cross-sections $\sigma(pp\to jj)$ at the LHC
with cuts ($p_T(j)>20\GeV$, $|\eta(j)|<5.0$, 
$\Delta R(j,j)>0.5$). 
}
\label{Tab:cs_jj}
\end{table}

\begin{figure}
\begin{center}
\centerline{
\epsfxsize=17.0cm
\epsfysize=17.0cm
\epsffile{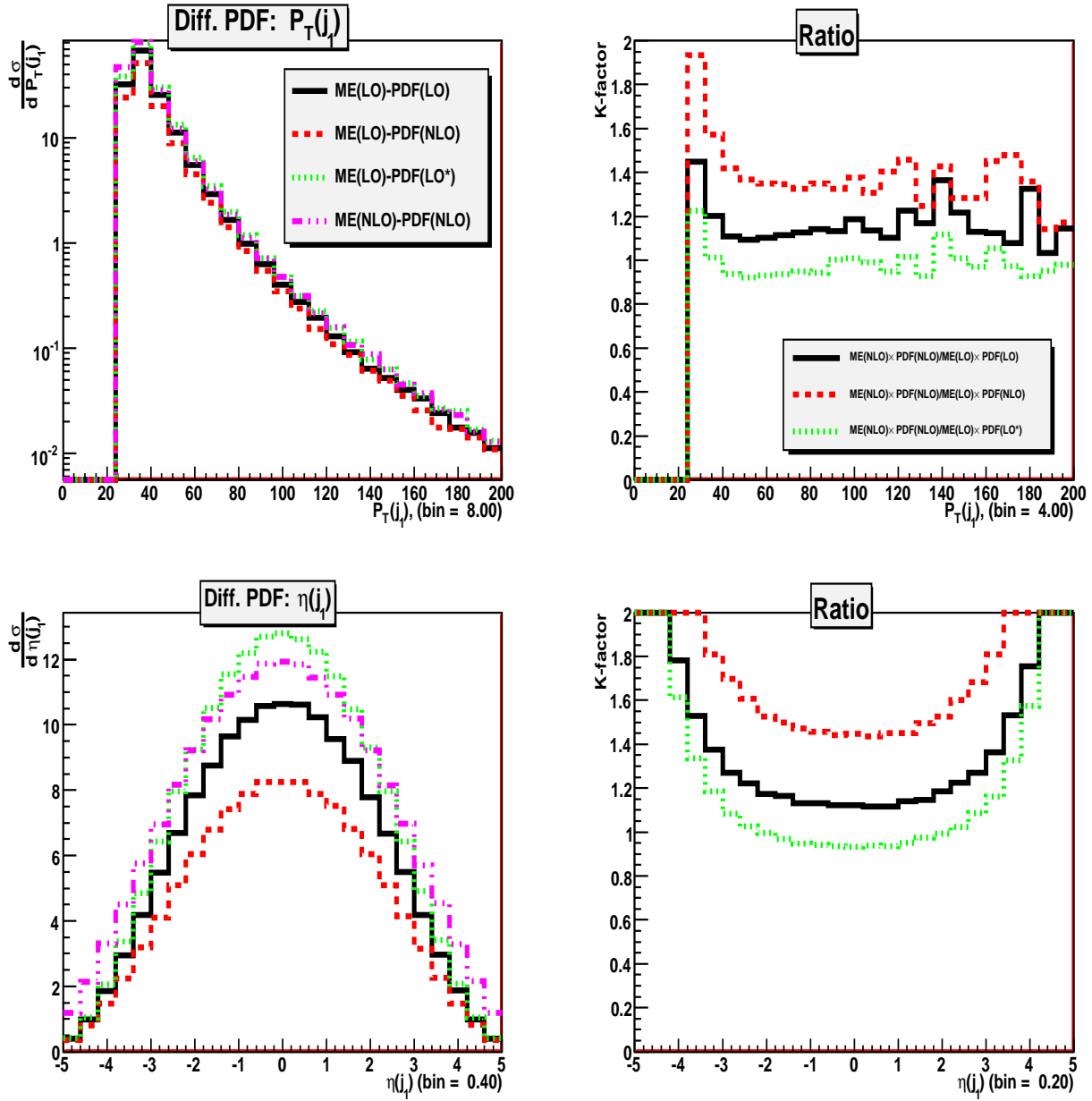}
}
\vspace{0.3cm}
\caption{
Inclusive di-jet cross-section at the LHC.
Differential cross-sections for the highest-Pt jet in the inclusive jet 
production at the LHC as a function of $p_T$ (upper plots) and for $\eta$ 
(lower plots).
}
\vspace{-1cm}
\label{Pic:jj_phys}
\end{center}
\end{figure}

As our final example we look at high-$E_T$ jet production at the LHC. In 
this case as we span the range of $E_T$ we span a range of $x$, i.e. at LO 
$x=E_T/\sqrt{s}$. We also change the dominant mechanism as we change $E_T$ --
at the highest $E_T$  quark-quark interactions 
are dominant while for $x<0.1$ gluon-gluon interactions take over. In the 
intermediate range there is also a large contribution from quark-gluon 
interactions. The jet production as a function of $E_T$ using the LO 
generator and the different pdfs as a ratio to the {\it truth} are shown for 
the lower $E_T$ values in 
Fig.~\ref{Pic:jj_phys}. For the lowest $p_T$, 
where we probe $x \sim 0.005$, the 
problems due to the relatively suppressed NLO gluon are again apparent and the 
LO and LO* pdfs are better in normalization and shape 
than NLO. Again the LO* pdfs provide the best description, very near to the
full NLO prediction as seen in table~\ref{Tab:cs_jj}. As we go to higher $E_T$
the predictions largely converge, but the NLO pdfs used in the LO calculation
tend to be a little small.   

\begin{figure}
\begin{center}
\centerline{
\epsfxsize=1.0
\textwidth\epsfbox{{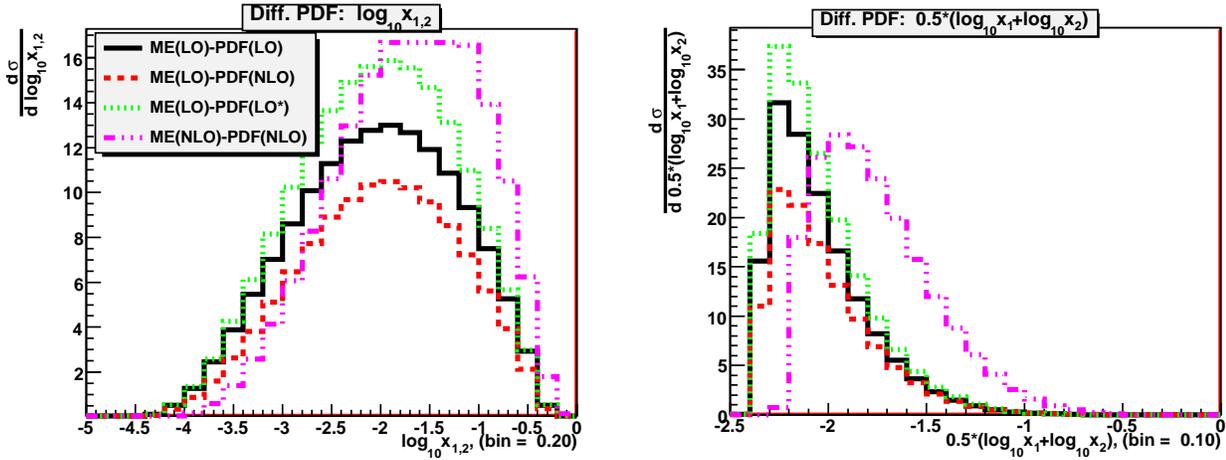}}
}
\vspace{0.3cm}
\caption{ 
The distributions of $x_{1,2}$ of the contributing parton distributions for 
the inclusive 2-jets production at the LHC in the different types of 
calculation. 
}
\label{Plot:jj_x}
\end{center}
\end{figure}

\section{Discussion}
There are a variety of reasons why the NLO cross-section corrections may be 
fairly large. As discussed earlier in the article, one major reason is when 
one probes small-$x$ parton distributions and there appears a $1/x$ 
divergence in the cross-section for the first time at NLO. This is 
particularly the case for gluon dominated processes, and is the main reason 
for the large corrections to the $\bb$ production at the LHC, but it
also contributes to the lowest $E_T$ jet cross-section and even probably a 
little to the Higgs cross-section from gluon-gluon fusion, 
since for the lowish mass used gluons in 
the region $x<0.001$ are probed. There are also large corrections due to 
soft gluon emission, which is important in regions near the edge of phase 
space, i.e. we have large so-called threshold corrections. These have been 
shown to have effects which persist some distance from the strict threshold 
region, and also contribute to the large Higgs production correction in 
gluon-gluon fusion. They are the dominant reason for the large correction in 
$t\bar t$ production. There can also be a large term appearing in NLO 
cross-sections from the analytic continuation from the space-like region where 
structure function coefficient functions are calculated to the time-like 
region of $s$-channel processes \cite{Parisi:1979xd}, 
and this contributes significantly to the large K-factors for $W$ and $Z$ production. 

Hence, even if one extracted the NLO parton distributions perfectly from 
comparison to mainly DIS data, using them along with LO cross-sections for 
processes at the LHC one would expect the predictions to be generally too 
low, often significantly. This is indeed seen to be the case. The particular 
problem at small $x$, where the NLO gluon distribution is far too small
to use with LO cross-sections, was the main motivation for introducing the 
LO* parton distributions, and it is no surprise that they give a far better 
prediction for the small-$x$ sensitive quantities, though the high quality of 
the quantitative prediction is surprising. However, the relaxation of the 
momentum sum rule has allowed the LO* gluon distribution to be bigger than 
that at NLO and standard LO at most values of $x$, and the quarks at $x<0.1$ 
driven by gluon evolution to follow suit. This actually results in the LO* 
parton distributions giving the best predictions for the quantities where NLO 
threshold  corrections are large, but where partons, particularly the gluon
distribution, are probed for $x<0.1$. In fact, since the parton distributions 
fall away very quickly above $x=0.1$ at high scales, the region where 
$x \sim 0.1$ is probed is effectively near threshold. Also, the fact that the 
increased LO* gluon allows the quarks to be bigger than even NLO for $x<0.01$
improves the prediction for the vector boson production (particularly 
when an additional jet is present), compensating for the NLO correction due 
to analytic continuation. Overall the LO* parton distributions can 
qualitatively mimic the effect of the NLO cross-section corrections for a 
surprisingly wide variety of processes. 

However, $t$-channel dominated processes, illustrated here by single top 
production and Higgs production via vector-boson fusion, generally 
do not correspond 
to any of the cases where the NLO correction is expected to be large. The 
term from analytic continuation is not present. Moreover, a sharp threshold 
is not so well defined in a $t$-channel process, so one does not see large 
threshold corrections in the same way. However, the fact that there is more 
activity in the final state in a $t$-channel process than a similar 
$s$-channel process means that the ``average'' $x$ probed is 
higher, e.g. nearly 
an order of magnitude for vector-boson production of Higgs compared to 
gluon-gluon fusion, and there is less likelihood of corrections due to 
small-$x$ divergences in the cross-section. Hence the absolute NLO 
cross-section correction is very small in these cases, the K-factor when NLO 
partons are used in both cases being very close to unity, and in fact 
slightly less. While the LO partons for the relevant $x \sim 0.1$ are smaller 
than at NLO, giving too small a prediction, the LO* partons are generally 
enhanced compared to NLO, and give a slightly too large prediction. Hence, 
for our examples of $t$-channel processes the NLO partons are generally the 
absolute best to use with LO cross-sections, but one is only off by at most 
$10\%$ in normalization, with no problem with shape, using either LO or LO* 
parton distributions.

\section{Conclusions}

We have examined the effects of varying both the order of the matrix elements 
and the parton distributions when calculating cross-section for particle 
colliders which involve hadrons. The intention is to find the best set of 
parton distributions to use if one only has LO matrix elements, as is often 
the case when using Monte Carlo generators. In order to do this we have 
calculated a variety of processes, both inclusive and more differential, 
where we can currently find the result using both NLO matrix elements
and parton distributions. In the case where we look at the details of the 
final state, we have used Monte Carlo generators with either LO or NLO matrix 
elements and LO parton showering. 

We notice that the NLO matrix element corrections 
are generally positive, and sometimes 
give very large enhancements. In contrast, the parton distributions at NLO
are sometimes bigger, but also sometimes smaller. Indeed, this must be the 
case since they satisfy sum rules on both parton number and momentum. 
The fact that higher order partonic matrix elements, and sometimes splitting 
functions contain large terms in $\ln(1-x)$ and $\ln(1/x)$
means that the partons extracted in these regions at LO tend to be enhanced to 
make up for these missing terms. From the sum rules 
the parton distributions away from these regions are correspondingly
generally smaller than at higher orders. In practice, 
there is a pronounced depletion of quarks for $x \sim 0.1-0.001$, and the 
large LO gluon at very small $x$ means it is smaller above $x\sim 0.01$.  
These are often the regions of most interest at hadron colliders.
 
A fixed prescription of either LO or NLO pdfs with LO matrix elements is 
unsuccessful, with each significantly out in some cases. For LO pdfs this is 
mainly due to the depletion just discussed, while for NLO partons the 
smallness 
in some regions compared to LO pdfs is a major problem if the large NLO matrix 
element is absent. This is particularly the case at very small $x$, where the 
large LO gluon compensates for the missing large NLO corrections. The best 
features of each order are required. 

To this end we have suggested an optimal set of partons for Monte Carlos, 
which is essentially LO but with modifications to make results more NLO-like, 
and are called LO* pdfs. The NLO coupling is used, which is larger at low 
scales, and helps give a good fit to the data used when extracting partons 
from a global fit. The momentum sum rule is also relaxed for the input parton 
distributions. This allows LO pdfs to be large where it is required for them 
to compensate for missing higher order corrections, but not correspondingly 
depleted elsewhere. 

We have compared the LO, NLO and LO* pdfs in LO calculations to the 
{\it truth}, i.e. full NLO, for a wide variety of processes which probe 
different types of pdf, ranges of $x$ and QCD scales. In general, the results 
are very positive. The LO* pdfs nearly always provide the best description 
compared 
to the {\it truth}, especially for the $s$-channel processes. This is 
particularly the case in terms of the normalization, but the shape is 
usually at least as good as -- and sometimes much better than -- when 
using NLO 
pdfs. It is noticeable that the type of enhancement of the pdfs provided by 
the allowed momentum violation seems to give a reasonably universally 
correct modification of the LO pdfs. Hence, while being rather crude, it 
seems to produce the effects obtained by modifying pdfs in a process 
dependent fashion for use in Monte Carlo generators/resummations discussed 
in e.g.~\cite{Mrenna:1999mq} (based on the calculation of the hard Wilson 
coefficient functions for the $Q_T$-divergent part of the cross-sections in 
the context of $Q_T$ resummation \cite{Davies:1984hs}). In particular it 
produces a relatively 
small enhancement of quarks and a rather larger enhancement of the gluon 
distribution. 

It should be noted that no modification of the pdfs can hope to reproduce 
successfully 
all the features of genuine NLO corrections. In particular we 
noticed 
the recurrent feature that the high-$p_T$ distributions are underestimated 
using the LO generators, and this can only be corrected by the inclusion of 
the emission of a relatively hard additional parton which occurs in the NLO 
matrix element correction. However, if one is limited to a strictly LO 
calculation, the shape as a function of $p_T$ seems generally fairly 
independent 
of the pdf, and the normalization is usually best with the LO* pdfs. 

A preliminary version of the LO* pdfs, based on fitting the same data as in 
\cite{Martin:2004ir}, is available on request from the authors. A more 
up-to-date 
version, based on a fit to all recent data, and with uncertainty bands for the 
pdfs, will be provided in future global fits, i.e. will follow similar lines 
to the MSTW07 NLO pdfs discussed in \cite{Thorne:2007bt}. 

\section{Acknowledgements}

We would like to thank Jon Butterworth for asking questions which started 
this project; Sven Moch for encouraging further investigation for the DIS07 
workshop and Claire Gwenlan for help during the early stages and for 
subsequent discussions and aid. We would also like to thank Paolo Bartalini, 
Mandy Cooper-Sarkar, Joey Huston, Alan Martin, Steve Mrenna, T. Sj\"ostrand, 
James Stirling, Graeme Watt and Bryan Webber for helpful discussions. 
RST would like to thank the Royal Society for the award of a University 
Research Fellowship. AS would like to thank the Science and Technology 
Facilities Council for the award of a Responsive Research Associate position, 
RFBR (the RFBR grant 07-07-00365-a) and the MCnet Marie Curie Research 
Training Network for partial support of the project. \\

\end{document}